\documentclass[10pt]{article}

\usepackage{amsmath}
\usepackage{amssymb}

\usepackage{graphicx}

\usepackage{cite}

\usepackage{color} 

\usepackage[labelfont=bf,labelsep=period,justification=raggedright]{caption}

\makeatletter
\renewcommand{\@biblabel}[1]{\quad#1.}
\makeatother

\date{}

\newcommand{\vks}{V\mathrm{_{K2}^{shift}}}
\newcommand{\gsyn}{g^\mathrm{syn}}
\newcommand{\gas}{g_\mathrm{\circlearrowright}}
\newcommand{\OneTwoThree}{\left (1 \! \prec \! 2 \! \prec \! 3 \right)}
\newcommand{\OneThreeTwo}{\left (1 \! \prec \! 3 \! \prec \! 2 \right)}
\newcommand{\OnePerpTwoThree}{ \left (1 \!  \perp \! \{2 \! \parallel \! 3\} \right)}
\newcommand{\TwoPerpOneThree}{ \left (2 \!  \perp \!  \{1\! \parallel \! 3\} \right )}
\newcommand{\ThreePerpOneTwo}{ \left (3 \!  \perp \!  \{1 \! \parallel \! 2\} \right)}
\begin{document}

% Title must be 150 characters or less
\begin{flushleft}
{\Large
\textbf{Key bifurcations of bursting polyrhythms in 3-cell central pattern generators}
}
% Insert Author names, affiliations and corresponding author email.
\\
Jeremy Wojcik$^{1}$,  Robert Clewley$^{1,2}$, Justus Schwabedal$^{2}$, and  Andrey L. Shilnikov$^{1,2,3,\ast}$
\\
\bf{1} Department of Mathematics and Statistics, Georgia State University, Atlanta, GA 30303, USA\\
\bf{2} Neuroscience Institute, Georgia State University, Atlanta, GA 30303, USA\\
\bf{3} Department of Computational Mathematics, Nizhny Novgorod State University, Nizhny Novgorod, Russia  \\
$\ast$ E-mail: ashilnikov@gsu.edu
\end{flushleft}

% Please keep the abstract between 250 and 300 words
\section*{Abstract}
We identify and describe the key qualitative rhythmic states in various 3-cell network motifs of a multifunctional central pattern generator (CPG). Such CPGs are neural microcircuits of cells whose synergetic interactions produce multiple states with distinct phase-locked patterns of bursting activity. To study biologically plausible CPG models, we develop  a suite of computational tools that reduce the problem of stability and existence of rhythmic patterns in networks to the bifurcation analysis of fixed points and invariant curves of a Poincar\'e return maps for phase lags between cells. 

We explore different functional possibilities for motifs involving symmetry breaking and heterogeneity. This is achieved by varying coupling properties of the synapses between the cells and studying the qualitative changes in the structure of the corresponding return maps. Our findings provide a systematic basis for understanding plausible biophysical mechanisms for the regulation of rhythmic patterns generated by various CPGs in the context of motor control such as gait-switching in locomotion. Our analysis does not require knowledge of the equations modeling the system and provides a powerful qualitative approach to studying detailed models of rhythmic behavior. Thus, our approach is applicable to a wide range of biological phenomena beyond motor control.

% Please keep the Author Summary between 150 and 200 words
% Use first person. PLoS ONE authors please skip this step. 
% Author Summary not valid for PLoS ONE submissions.   
\section*{Author Summary}
Rhythmic motor behaviors, such as heartbeat, respiration, chewing, and
locomotion are often independently produced by small networks of neurons called Central
Pattern Generators (CPGs). It is not clear either what mechanisms a single
motor system can use to generate multiple rhythms, for instance: whether CPGs use
dedicated circuitry for each function or whether the same circuitry is
can govern several behaviors. A systematic way to explore
this is to create mathematical models that use biologically plausible components
and classify the possible varieties of rhythmic outcomes. 

We performed such a study of CPG networks based on three inter-connected neurons. We systematically varied the connection types and strengths between the neurons to discover how these affect the behavioral repertoire of the CPG. To do this, we created a geometric representation of the simulated CPG behavior of each possible configuration of the network, which greatly simplifies the study of the 9-dimensional system of nonlinear differential equations. We discovered several configurations that support multiple rhythms and characterized their robustness. By varying physiologically reasonable parameters in the model, we also describe some mechanisms by which a biological system could be switched between its multiple stable rhythmic states.

%%%%%%%%%%%%%%%%%%%%%%%%%%%%%%%%%%%%%%%%%%%%%%%%%%%%%%%%%%%%%%%%%%%%%%%%%%%%%%%%%%%%%%%%%%%%%%%%%%%%%%%%%%%%%%%%

\section*{Introduction}

A central pattern generator (CPG) is a neural circuit of cells whose synergetic interactions can autonomously  produce rhythmic patterns of activity that determine  vital motor behaviors in animals \cite{CPG,Bal1988,Marder1996}. CPGs have been identified and studied in many animals, where they have been implicated in the control of diverse behaviors such as heartbeat, sleep, respiration, chewing, and locomotion on land and in water \cite{KCF05,CTMKF07,SNLK11,NSLGK012}. 
Mathematical modeling studies, at both abstract and realistic levels of description, have provided useful insights into the operational principles of CPGs\cite{Kopell26102004,Ma87,Kopell88,Canavier1994,SKM94,Dror1999,Prinz2003}. Although many dynamic models of specific CPGs have been developed, it remains unclear how CPGs achieve the level of robustness and stability observed in nature \cite{Best2005,prl08,Shilnikov2008b,SHWG10, Koch01122011, Wojcik2011a, Calabrese01122011,M12}.

A common component of many identified CPGs is a half-center oscillator (HCO), which is composed of two bilaterally symmetric cells reciprocally inhibiting each other to produce an alternating anti-phase bursting pattern \cite{Brown08121911}.  There has been much work on the mechanisms giving rise to anti-phase bursting in relaxation HCO networks, including synaptic release, escape and post-inhibitory rebound \cite{Rubin2000a,SZVC99}. Studies of HCOs composed of Hodgkin-Huxley type model cells have also demonstrated the possibility of bistability and the coexistence of several in-phase and anti-phase bursting patterns based on synaptic time scales or delays \cite{vae94,pre2010,pre2012}.

We are interested in exploring the constituent building blocks --- or ``motifs'' --- that make up more complex CPG circuits, and the dynamic principles behind the support of more general multi-stable rhythmic patterns \cite{Dror1999, Wojcik2011a,TermanRubin2012}. We will refer to multi-stable rhythmic patterns as ``polyrhythms.'' We consider the range of basic motifs comprising of three biophysical neurons and their chemical synapses, and how those relate to, and can be understood from the known principles of two-cell HCOs. We will study the roles of asymmetric and unique connections, and the intrinsic properties of their associated neurons, in generating a set of coexisting synchronous patterns of bursting waveforms. The particular kinds of network structure that we study here reflect the known physiology of various CPG networks in real animals. Many anatomically and physiologically diverse CPG circuits involve a three-cell motif \cite{Milo25102002,Sporns2004}, including the spiny lobster pyloric network \cite{CPG,RevModPhys.78.1213}, the \emph{Tritonia} swim circuit, and the \emph{Lymnaea} respiratory CPGs \cite{Bulloch1992422,Marder1994752,Frost09011996,KatzHooper07}.

An important open question in the experimental study of real CPGs is whether they use dedicated circuitry for each output pattern, or whether the same circuitry is multi-functional \cite{Kristan2008b,Briggman2008}, i.e. can govern several behaviors.  Switching between multi-stable rhythms can be attributed to input-dependent switching between attractors of the CPG, where each attractor is associated with a specific rhythm. Our goal is to characterize how observed multi-stable states arise from the coupling, and also to suggest how real circuits may take advantage of the multi-stable states to dynamically switch between rhythmic outputs. For example, we will show how motif rhythms are selected by changing the relative timing of bursts by physiologically plausible perturbations. We will also demonstrate how the set of possible rhythmic outcomes can be controlled by varying the duty cycle of bursts, and by varying the network coupling both symmetrically and asymmetrically \cite{Shilnikov2008b,Wojcik2011a}. We also consider the role of a small number of excitatory or electrical connections in an otherwise inhibitory network. Our greater goal is to gain insight into the rules governing pattern formation in complex networks of neurons, for which we believe one should first investigate the rules underlying the emergence of cooperative rhythms in smaller network motifs. 

In this work, we apply a novel computational tool that reduces the problem of stability and existence of bursting rhythms in large networks to the bifurcation analysis of fixed points (abbreviated FPs) and invariant circles of Poincar\'e return maps measuring the phase lags between the burst initiations in the cells.  
The structure of the phase space of the map reflects the characteristics the state space of the corresponding CPG motif. Equipped with the maps, we are able to predict and identify the set of robust bursting outcomes of the CPG. These states are either phase-locked or periodically varying lags corresponding to FP or invariant circle attractors (respectively) of the map.  Comprehensive simulations of the transient phasic relationships in the network are based on the delayed release of cells from a suppressed, hyperpolarized state. This complements the phase resetting technique allowing for thorough exploration of network dynamics with spiking cells \cite{PRC}. We demonstrate that more general synaptically-coupled networks possess stable bursting patterns that do not occur in similar three-oscillator motifs with gap junction coupling, which is bidirectionally symmetric \cite{Belykh05}.

%%%%%%%%%%

%%%%% FIGURE 1 around here
%\vspace{0.2cm}{\it FIGURE 1 around here} \vspace{0.2cm}

\begin{figure}[h!]
\begin{center}
\includegraphics[width=0.99  \textwidth]{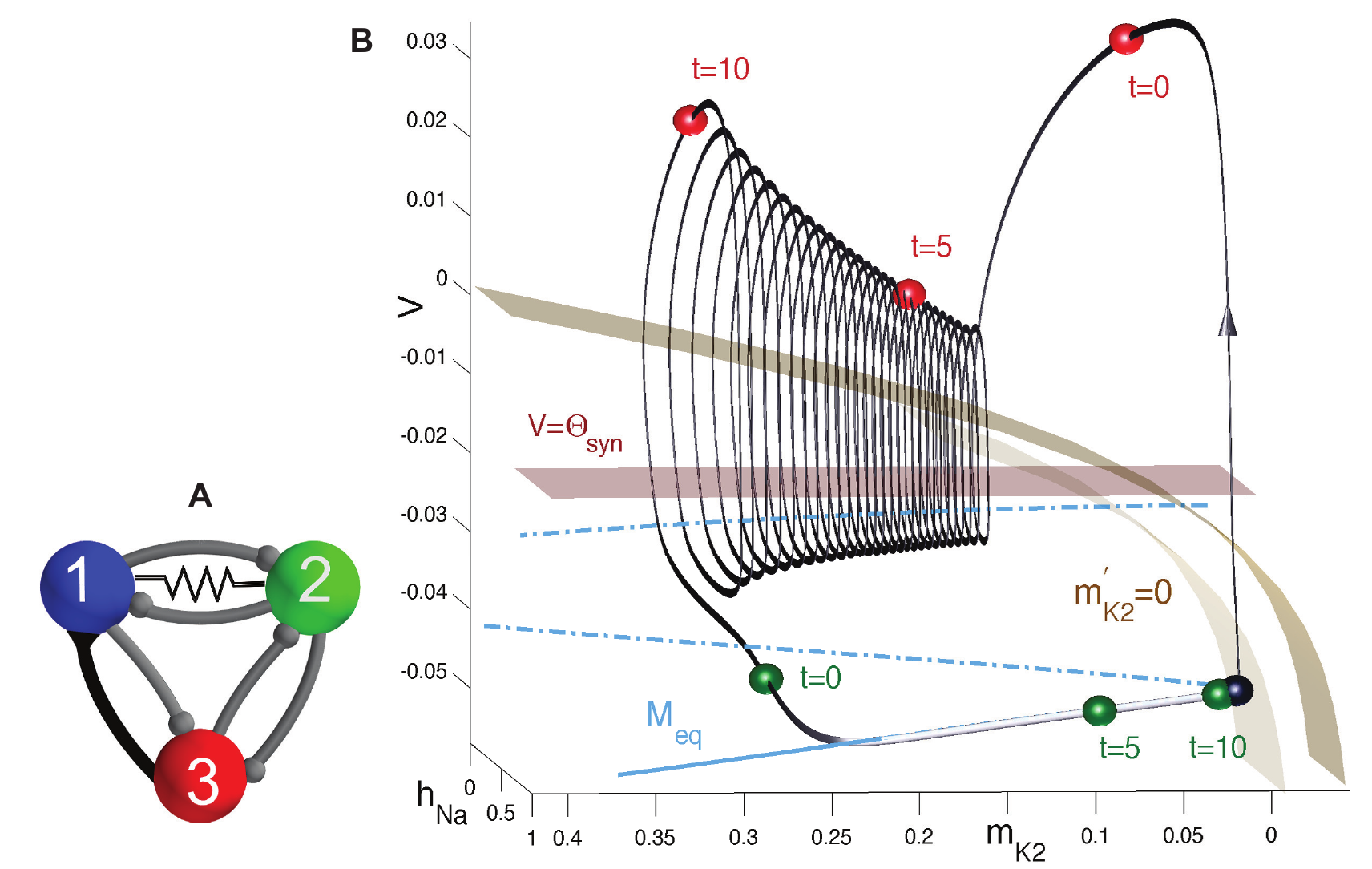}
\end{center}
\caption {{\bf Motif network diagram and phase space of typical bursting trajectory of single cell.} (A) Caricature of a mixed 3-cell motif with inhibitory and excitatory synapses, represented by $\bullet$ and $ \bigtriangledown$, resp., as well as an electrical connection through the gap junction between interneurons 1 and 2. (B) Bursting trajectory (gray) in the 3D phase space of the model, which is made of the ``active'' spiking (solenoid-like shaped) and the flat hyperpolarized sections.  
The gap between the 2D slow nullcline, $m^\prime_{\mathrm{K2}}=0$, and the low knee on the slow quiescent manifold, $\mathrm{M_{eq}}$, determines the amount of inhibition needed by the active pre-synaptic cell above the synaptic threshold, $\Theta_{\rm syn}$, to slow or keep the post-synaptic cell(s) at the hyperpolarized quiescent state around $-60\mathrm{mV}$. The positions of the red, green and blue spheres on the  bursting trajectory depict the phases of the weakly-connected cells of the CPG at two instances:  the active red cell inhibits, in anti-phase, the temporarily inactive green and blue cells at two time instances.}
\label{fig1}
\end{figure}

%%%% RESULTS

\section*{Results}

Our results are organized as follows: first, we introduce our new computational tools, which are based on 2D return maps for phase lags between oscillators.  We then present maps for symmetric inhibitory motifs and examine the structure of the maps depends on the duty cycle of bursting, i.e. on how close the individual neurons are to the boundaries between activity types (hyperpolarized quiescence and tonic spiking). Here, we also examine bifurcations that the map undergoes as the rotational symmetry of the reciprocally 3-cell motif is broken. This is followed by a detailed analysis of bifurcations of fixed point (FP) and invariant circle attractors of the maps for several characteristic configurations of asymmetric motifs, including a CPG based on a model of the pyloric circuit of a crustacean.  We conclude the inhibitory cases with the consideration of the fine structure of a map near a synchronous state. We then discuss the maps for 3-cell motifs with only excitatory synapses, which is followed by the examination of mixed inhibitory-excitatory motifs, and finally an inhibitory motif with an additional electrical synapse in the form of a gap junction.

%%%%%%%%%%%%%%%%%%%%%%%%%%%%%%%%%%%

\subsection*{A computational method for phase lag return mappings}

We introduce a new method for computationally analyzing phase-locked dynamics of our networks. This is a non-standard method that has general utility outside of  our application, and we therefore present it here as a scientific result.

We first introduce the types of trajectories we focus on and how we measure them. The reduced leech heart interneuron can demonstrate many regular and irregular activity types, including hyper- and de-polarized quiescence, tonic spiking and bursting oscillations. We focus on periodic bursting, and Figure~\ref{fig1} shows a  trajectory (dark gray) in the 3D phase space of the model. The helical coils of the trajectory correspond to the active tonic spiking period of bursting due to the fast sodium current. The flat section corresponds to the hyperpolarized quiescent portion of bursting due to the slow recovery of the potassium current. In Fig.~\ref{fig1}, two snapshots (at $t=0$ and $t=10 \; \mathrm{ms}$) depict the positions of the blue, green and red spheres representing the momentarily states of all three interneurons. The coupling between the cells is chosen weak so that  network interactions should only affect the relative phases of the cells on the intact bursting trajectory, i.e. without deforming the trajectory.

$\vks$ is a model parameter that measures the deviation from the half-activation voltage $V_{1/2}=-0.018\mathrm{V}$ of the potassium channel, $m^{\infty}_{\mathrm{K2}}=1/2$. We use $\vks$
as a bifurcation parameter to control the duty cycle (DC) of the interneurons.
The duty cycle is the fraction of the burst period in which the cell is spiking, and is a property known to affect the synchronization properties of coupled bursters \cite{Shilnikov2008b,prl08}. The individual cell remains bursting within the interval
$\vks \in [-0.024235,\,-0.01862]$. At smaller values of $\vks$, it begins oscillating tonically about the depolarized steady state, and becomes quiescent at greater values of $\vks$.
Therefore, the closer the cell is to either boundary, the DC of bursting becomes longer or shorter respective: the DC is about $80\%$ at $\vks =-0.0225\mathrm{V}$ and $25\%$ at $\vks =-0.01895\mathrm{V}$. For $50\%$ DC we set $\vks =-0.021\mathrm{V}$, in the middle of the bursting interval (see Fig.~\ref{fig3}).

%%%%% FIG 2 around here
%\vspace{0.2cm}{\it FIGURE 2 around here} \vspace{0.2cm}

\begin{figure}[h!]
\begin{center}
\includegraphics[width=0.75  \textwidth]{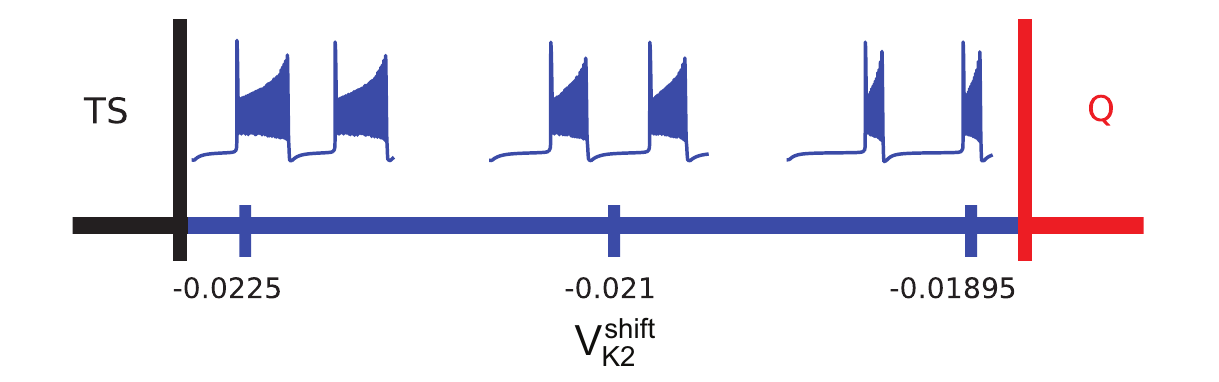}
\end{center}
\caption{ {\bf Schematic showing regimes and how burst duration changes as the bifurcation parameter, $\vks$ is varied.} Burst duration increases as $\vks$ approaches the boundary of the tonic spiking (TS) state, and decreases towards the boundary of hyperpolarized quiescence (Q). The post-synaptic cell on the network can temporarily cross either boundary when excited or inhibited by synaptic currents from pre-synaptic neurons.}
\label{fig2}
\end{figure}

When an isolated bursting cell is set close to a transition to either tonic spiking or hyperpolarized quiescence, its network dynamics become sensitive to external perturbations from its pre-synaptic cells. For example, when the post-synaptic cell is close to the tonic-spiking boundary, excitation can cause the post-synaptic cell to burst  longer or even move it (temporarily) over the boundary into the tonic spiking (TS) region. In contrast, inhibition shortens the duty cycle of the post-synaptic neuron if it does not completely suppress its activity (Fig.~\ref{fig2}).

%%%%% FIG 3 around here
%\vspace{0.2cm}{\it FIGURE 3 around here} \vspace{0.2cm}

\begin{figure}[h!]
\begin{center}
\includegraphics[width=0.99  \textwidth]{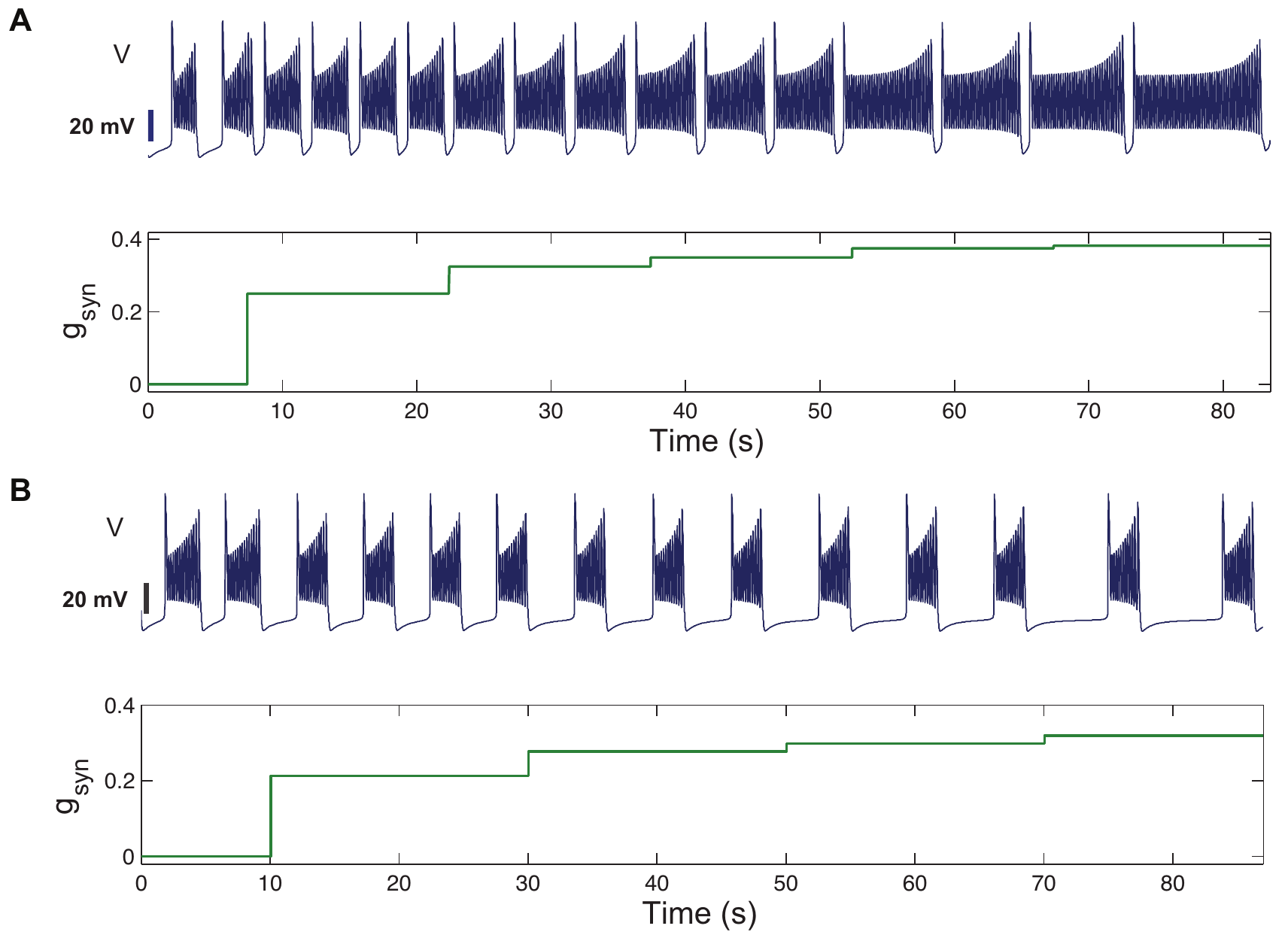}
\end{center}
\caption{ {\bf Variations of bursting of the post-synaptic cell with synaptic strength.} Step-wise increases in excitatory (top)  and inhibitory (bottom) strengths, $\gsyn$, from pre-synaptic cell(s). Increase of the duty cycle (DC) of bursting is  through  the extension of either the active phases of bursting  or the interburst intervals as  the post-synaptic cell on the network is shifted by synaptic perturbations toward either the tonic spiking (TS) or hyperpolarized quiescence (Q) boundaries in Fig.~\ref{fig2}.}
\label{fig3}
\end{figure}

\subsubsection*{The return map for phase lags}

We reduce the problem of the existence and stability of bursting rhythmic patterns to the bifurcation analysis of FPs and invariant curves of Poincar\'e return maps for phase lags between the neurons. In this study, we mostly consider relatively weakly coupled motifs, but our approach has no inherent limitation to weak coupling. Here, the weakly coupled case is a pilot study that lets us test our technique and also uncover all rhythms, both stable and unstable, that can possibly occur in the network. Detailed scrutiny of the return maps is computationally expensive: an exploration of one parameter set can take up to three three hours on a state-of-the-art desktop workstation depending on the accuracy of the mesh of initial conditions and length of the transients computed. 

%%%%% FIG 4 around here
%\vspace{0.2cm}{\it FIGURE 4 around here} \vspace{0.2cm}

\begin{figure}[h!]
\begin{center}
\includegraphics[width=0.6  \textwidth]{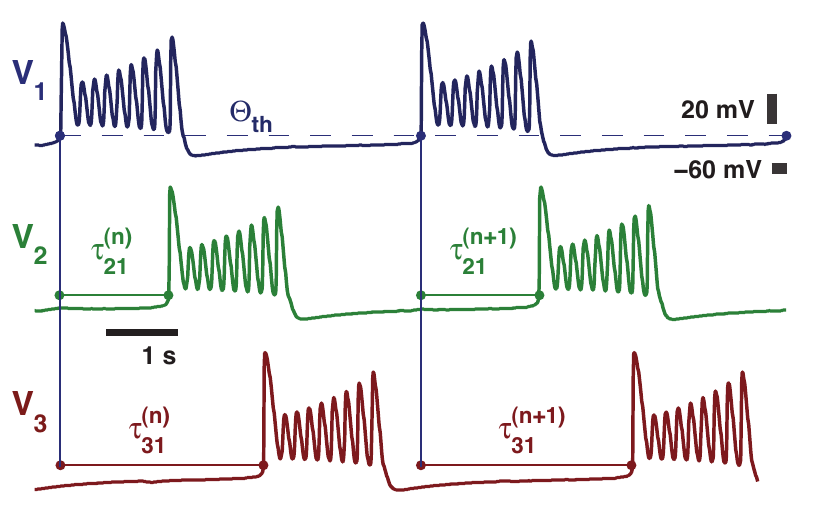}
\end{center}
\caption{ {\bf Sample voltage traces depicting phase measurements.} The phase of the reference cell~1 (blue) is reset when $V_1$ reaches an auxiliary threshold, $\Theta_\mathrm{th}=-40$ mV, at $\tau_1^{(n)}$. The recurrent time delays, $\tau_{21}^{(n)}$ and  $\tau_{31}^{(n)}$ between the burst onsets in cell~1 and cells 2 (green) and 3 (red), normalized over the cycle period, $\left [ \tau_1^{(n+1)}-\tau_1^{(n)} \right ] $,  define a sequence of phase lags: $\left \{ \Delta \phi_{21}^{(n)}, \Delta \phi_{31}^{(n)}  \right \}$.}
\label{fig4}
\end{figure}

The phase relationships between the coupled cells are defined through specific events,
 $\left \{ \tau^{(n)}_1,\, \tau^{(n)}_2,\, \tau^{(n)}_3  \right \}$, when their voltages cross a threshold, $\Theta_{th}$, from below. Such an event indicates the initiation of the $n^{\mathrm{th}}$ burst in the cells, see Fig.~\ref{fig4}.  We choose $\Theta_{th}=-0.04\mathrm{V}$, above the hyperpolarized voltage and below the spike oscillations within bursts.

We define a sequence of \emph{phase lags} by the delays in burst initiations relative to that of the reference cell~1, normalized over the current network period or the burst recurrent times for the reference cell, as follows:
\begin{equation}
\Delta \phi^{(n)} _{21} = \displaystyle{\frac{\tau^{(n+1)}_{21} - \tau^{(n)}_{21}}{\tau^{(n+1)}_{1} -\tau^{(n)} _{1}}}
\quad \mbox{and} \quad
\Delta \phi^{(n)} _{31} = \displaystyle{\frac{\tau^{(n+1)}_{31} - \tau^{(n)}_{31}}{\tau^{(n+1)}_{1} -\tau^{(n)} _{1}}}, \qquad \mbox{mod 1}.
\label{eq_lag}
\end{equation}

%%%%% FIGURE 5 around here
%\vspace{0.2cm}{\it FIGURE 5 around here} \vspace{0.2cm}
\begin{figure}[h!]
\begin{center}
\includegraphics[width=0.8  \textwidth]{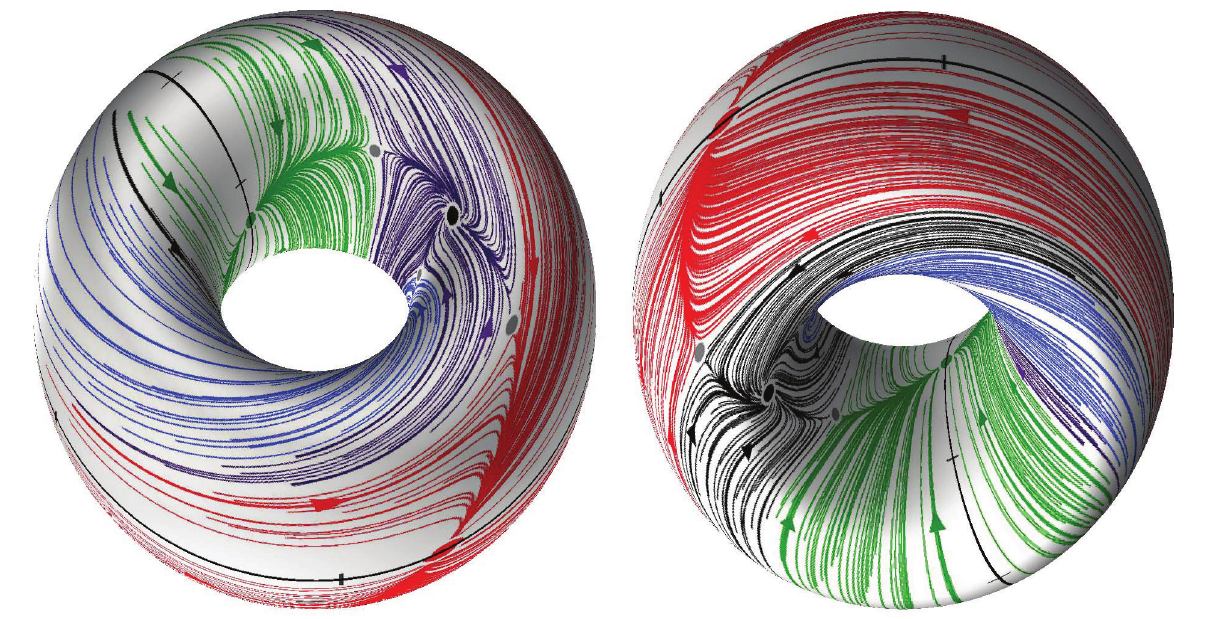}
\end{center}
\caption{ {\bf Poincar\'e return maps defined on the torus.} The return maps for the phase lags $\{ \Delta \phi_{21}^{(n)}, \Delta \phi_{31}^{(n)} \}$ between homogeneous cells at $50\%$ DC correspond to trajectories on a 2D torus $[0,1)\times [0,1)$. Different colors denote attractor basins of several FPs corresponding to phase locked states of distinct bursting rhythms.}
\label{fig5}
\end{figure}

An ordered pair, ${\mathbf M}_{n} = \left (\Delta \phi^ {(n)} _{21}, \Delta \phi^{(n)} _{31} \right ) $, defines a forward iterate, or a phase point, of the Poincar\'e return map for the phase lags:
\begin{equation}\label{eq_map}
\Pi :~ {\mathbf M}_n  \to {\mathbf M}_{n+1}
\end{equation}
A sequence, $\left \{(\Delta \phi^{(n)} _{21}, \Delta \phi^{(n)} _{31}) \right \}_{n=0}^N$, yields a forward
\emph{phase lag trajectory}, $\left \{ \mathbf M_{n} \right \}_{n=0}^N$, of the Poincar\'e return map on a 2D torus  $[0,1)\times [0,1)$ with phases defined on mod~1 (Fig.~\ref{fig5}).
Typically, such a trajectory is run for $N=90$ bursting cycles in our simulations. The run can be stopped when  the distance between several successive iterates becomes less than some preset value, say $|| \mathbf M_n - \mathbf M_{n+k} || < 10^{-3}$ and $k=5$. This is taken to mean that the trajectory has converged to a fixed point, $\mathbf M^*$, of the map. This FP corresponds to a phase locked rhythm and its coordinates correspond to specific constant phase lags between the cells. By varying the initial delays between cells~2 and 3 with respect to the reference cell~1,  we can detect any and all FPs of the map and identify the corresponding attractor basins and their boundaries.

We say that coupling is \emph{weak} between two cells of a motif when the convergence rate to any stable FP of the return map is slow. This means that the distance between any two successive iterates of a trajectory of the return map remains smaller than some bound, e.g.  $  \max || \mathbf M_n - \mathbf M_{n+1}|| < 0.05$. Therefore, we can say that coupling is relatively strong if a remote transient reaches a FP of the map  after just a few iterates. We point out that the  convergence can be quick even for nominally small $\gsyn$  provided that an individual cell is sufficiently close to either boundary of bursting activity (tonic spiking or quiescence).   

We conclude this section with some technical remarks concerning computational derivations of the map, $\Pi$. \emph{A priori}, the initial period (recurrence time) of the motif's dynamics is unknown due to the unknown outcome of nonlinear cell interactions; furthermore, it varies over the course of the bursting transient until it converges to a fixed value on the phase locked state.  We can estimate the initial phase lag in a first order approximation,
$\left (\Delta \phi^{(0)} _{21},\Delta \phi^{(0)} _{31} \right )$ between the networked neurons,
as the phase lags $ \left (\Delta \phi ^{\star}_{21}, \Delta \phi ^{\star} _{31} \right)$ on the periodic synchronous solution
of period $T_{synch}$. Note that $\Delta \phi ^{\star}_{21}$ is shifted away from $\Delta \phi ^{\star} _{31}$, i.e. is advanced or delayed.
Notice that, in the  weakly coupled case, the recurrent times of the reference cell are close to $T_{synch}$, which implies
 $(\Delta \phi ^{\star}_{21}, \Delta \phi ^{\star} _{31}) \approx (\Delta \phi ^{(0)}_{21}, \Delta \phi ^{(0)} _{31})$.
By setting $\Delta \phi_{21}=\Delta \phi_{31}=0$ and $t_{1j}=0$ at $V_1=\Theta_{th}$ we can parameterize the synchronous solution by a time shift, $\left \{ 0 \leq \tau_{1j} < T_{synch} \right \}$ or, alternatively, by phase lags $\left \{0 \leq \Delta \phi_{j1} < 1 \right \}$.
Thus, we can determine the initial phase lags by releasing the reference cell
 and having the others suppressed by external inhibitory pulses for $\tau_{12}=\Delta \phi^{(0)} _{21} T_{synch}$ and  $\tau_{13}=\Delta \phi^{(0)} _{31} T_{synch}$ from the same initial point, given by  $V_{i}=\Theta_{th}=-0.04\mathrm{V}$, on the synchronous
bursting trajectory. In essence, this means that initial phases between the reference cell and another can be set by releasing the latter from inhibition at various delays.

To complete a single phase lag map we choose the initial phase lags to be uniformly distributed on a grid of at least $40 \times 40$ points over the $[0,1) \times [0,1)$ torus. The initial guess for the phase lag distribution is based on knowledge of a trajectory for a synchronized motif. This guess will therefore differ from the self-consistent phase lag distribution once the networked cells begin to interact, especially with coupling strength variations. Similarly, the estimated network period, $T_{synch}$, will differ from the network's actual self-consistent period. In computations, this may result in fast jumps from the set of guessed initial phases from $n=0$ to $n=1$. These jumps are artifacts of our setup and not relevant to our study of the attractors, and so we begin recording the phase lag trajectory settled from the second bursting cycle. Due to weak coupling, transients do not evolve quickly, and we connect phase lag iterates of the map by straight lines in order to demonstrate and preserve the forward order, making them superficially resemble continuous-time vector fields in a plane. Lastly, we unfold the torus onto a unit square for the sake of visibility.

%%%%%
\subsection*{Multistability and duty cycle in homogenous inhibitory motifs} \label{inhib}

We first examine three homogeneous (permutationally symmetric) configurations of the network with nearly identical cells and connections. We demonstrate that these symmetric network motifs are multistable and hence able to produce several  coexisting bursting patterns. The homogeneous case allows us to reveal the role of the duty cycle as an order parameter that determines what robust patterns are observable. We suggest a biologically plausible  switching mechanism between the possible bursting patterns by application of a small hyperpolarized current that temporarily blocks  a targeted cell.

%%%%%%%%%
\subsubsection*{Short length motif}

We begin with a weakly coupled with $\gsyn=5 \times 10^{-4}$, homogeneous motif with $25\%$ DC and $\vks =-0.01895\mathrm{V}$, which is close to
the transition boundary between bursting and hyperpolarized quiescence. The proximity to the boundary means that even weak inhibition is able to suppress a postsynaptic cell that is near the hyperpolarized quiescent state (Figs.~\ref{fig1} and \ref{fig3}).

Figure~\ref{fig5}A shows the transient behaviors of the iterates of the phase lags $\Delta \phi_{21}^{(n)}$ and $\Delta \phi_{31}^{(n)}$ (shown in blue and gray colors) arising from initial conditions distributed uniformly over the unit interval.  The phase lags exponentially converge to phase-locked states near $0$ and $\frac{1}{2}$.

%%% FIGURE 6 ABOUT HERE
%\vspace{0.2cm}{\it FIGURE 6 around here} \vspace{0.2cm}

\begin{figure}[h!]
\begin{center}
\includegraphics[width=0.99  \textwidth]{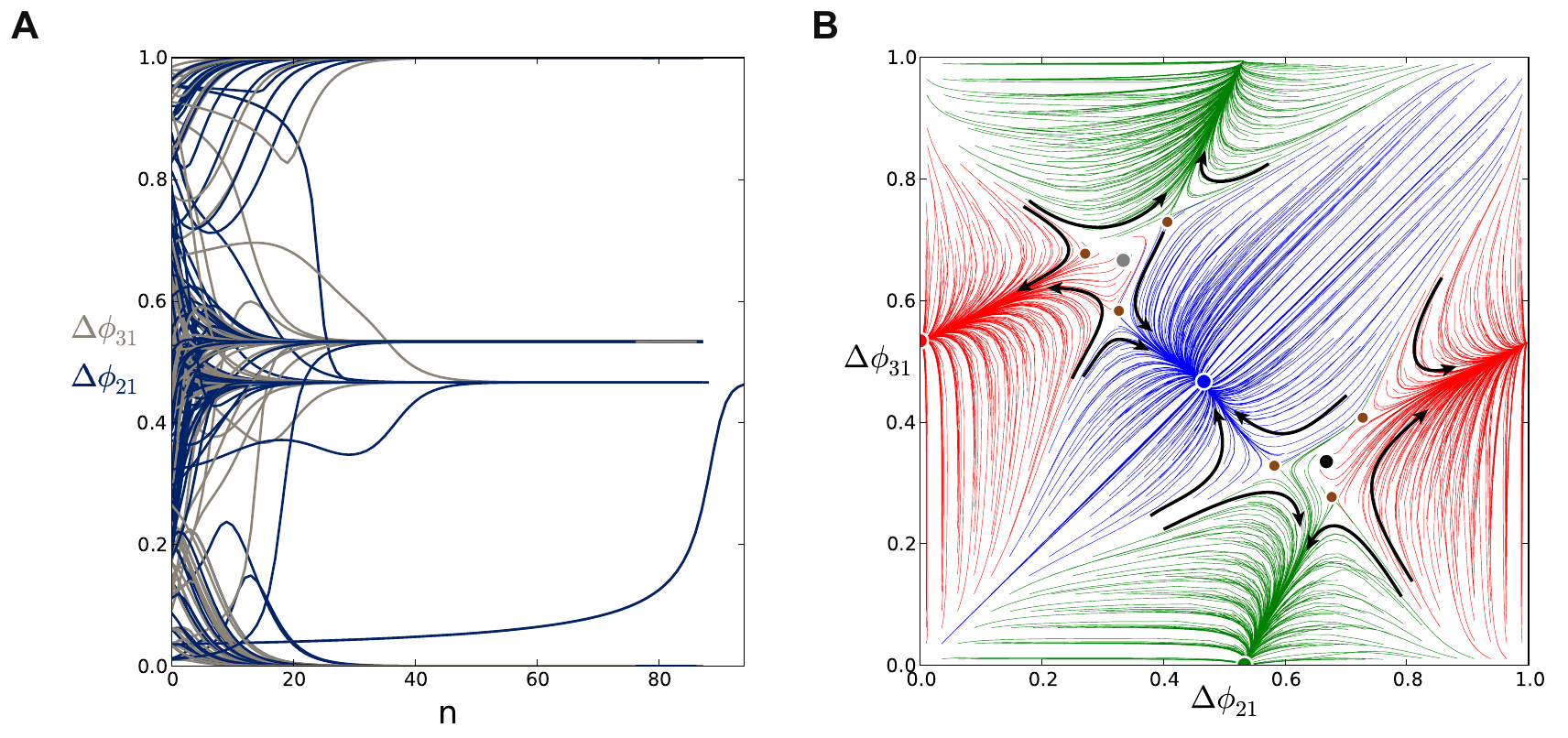}
\end{center}
\caption{ {\bf A comparison of time evolutions of phase lags and their motion in the 2D space of phase differences.} (A) Time evolutions of the phase lags, $\Delta \phi_{31}$ (gray) and $\Delta \phi_{21}$ (blue), exponentially converging to phase locked states after 50 burst cycles with short duty cycle, $\gsyn=5 \times 10^{-4}$. (B)
The corresponding Poincar\'e phase lag map revealing  three stable FPs  (shown in blue, red and green) at  
$ \left (\Delta \phi_{21},\, \Delta \phi_{31} \right ) = \left (\frac{1}{2},\,\frac{1}{2} \right )$, $ \left (0,\,\frac{1}{2} \right)$, $ \left (\frac{1}{2},\,0 \right )$
and two unstable FPs (dark dots) at $\left (\frac{2}{3},\frac{1}{3} \right )$ and $\left( \frac{1}{3},\frac{2}{3} \right )$. The attractor basins of three stable FPs are separated by the separatrices of six saddle FPs (smaller dots). Arrows on representative lines that connect iterates indicate the forward direction of iterates.}
\label{fig6}
\end{figure}

Using Eq.~(\ref{eq_map}), we compute the map $\Pi$ that is shown in Fig.~\ref{fig6}B. The projection of the map onto the unit square is an efficient way to represent the synchronized evolution of the phase lags and facilitates easy identification of the phase-locked states. These states are identified by three coexisting stable FPs or attractors of the system to which all forward iterates
converge. Here, the FPs are: red at $ \left ( \Delta \phi_{21} \approx 0, \,  \Delta \phi_{31} \approx \frac{1}{2} \right )$, green at $\left (\frac{1}{2},0 \right )$, and blue at $ \left ( \frac{1}{2},\frac{1}{2} \right) $. The attractor
basins of the stable FPs are shown in the corresponding colors. The attractor basins are subdivided by separatrices (incoming and outgoing sets) of six saddle FPs (shown by small dots) in the map. The robustness of a rhythm to perturbations is related to the size of its attractor basin. Similarly, FPs that have much larger basins than others can be thought of as ``dominating'' the phase plane in terms of likelihood of becoming the active state for a random initial condition or perturbation. Two triplets of saddles surround two more
unstable FPs located at approximately  $\left (\frac{2}{3},\frac{1}{3} \right )$ and  $\left ( \frac{1}{3},\frac{2}{3} \right )$.
The immediate neighborhood of the origin has a complex structure at high magnification, but globally it acts as a repeller (see the corresponding section on the fine structure near the origin).

%%%%% FIG 7 around here
%\vspace{0.2cm}{\it FIGURE 7 around here} \vspace{0.2cm}

\begin{figure}[h!]
\begin{center}
\includegraphics[width=0.99  \textwidth]{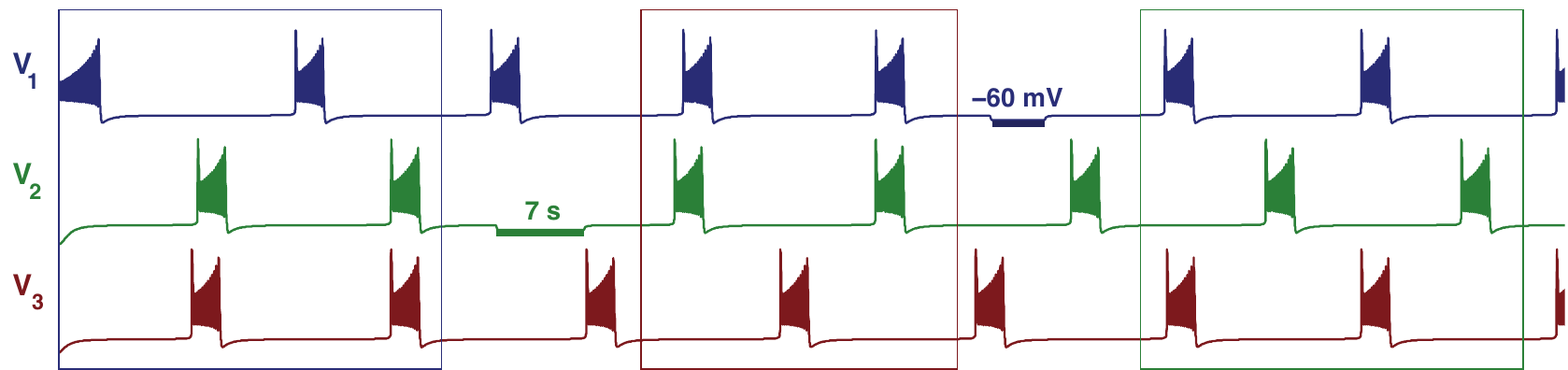}
\end{center}
\caption{ {\bf Time evolutions of voltage traces in the short bursting motif showing switching between coexisting rhythms.} Three coexisting stable rhythms:  $\OnePerpTwoThree$ (first episode), $\ThreePerpOneTwo$ (second episode) and
$\TwoPerpOneThree$ (third episode) in the short bursting motif with $25\%$ DC with $\pm 5\%$ random perturbations
applied to all inhibitory connections with $\gsyn=0.0005$.  Switching between rhythms is achieved by the application of appropriately-timed hyperpolarized pulses that release the targeted cells.}
\label{fig7}
\end{figure}

Let us interpret the role of a stable FP, for example the red one, in terms of phase-locked bursting patterns. Since the phases are defined modulo one, the coordinates $(\Delta \phi_{21}, \, \Delta \phi_{31} ) = (0 , \, \frac{1}{2})$, imply
that the corresponding rhythm is characterized by two fixed conditions $\phi_{1}=\phi_{2}$ and $\phi_{1}-\phi_{3}=\frac{1}{2}$. In other words, the reference cell fires in-phase (synchronized) with cell~2 and in anti-phase (anti-synchronized) with cell~3. Symbolically, we will use the following notation for this rhythm: $\ThreePerpOneTwo$,  in which in-phase and anti-phase bursting are represented by ($\Delta \phi_{12}=0$, or $\parallel$) and ($\Delta \phi_{13} = \frac{1}{2}$ or $\perp$), respectively.

Following this notation, the stable FP (blue) at $ \left ( \frac{1}{2},\frac{1}{2} \right )$ corresponds to the robust $\OnePerpTwoThree$ pattern, while the stable (green) FP $ \left ( \frac{1}{2}, \, 0 \right ) $ corresponds to the $\TwoPerpOneThree$ pattern. These coexisting
bursting rhythms are shown in Fig.~\ref{fig7}.
The motif can be made to switch between the polyrhythms by applying external pulses of appropriate duration to the targeted cells.

%%%%%%%% FIGURE 8 ABOUT HERE
%\vspace{0.2cm}{\it FIGURE 8 around here} \vspace{0.2cm}

\begin{figure}[h!]
\begin{center}
\includegraphics[width=0.99  \textwidth]{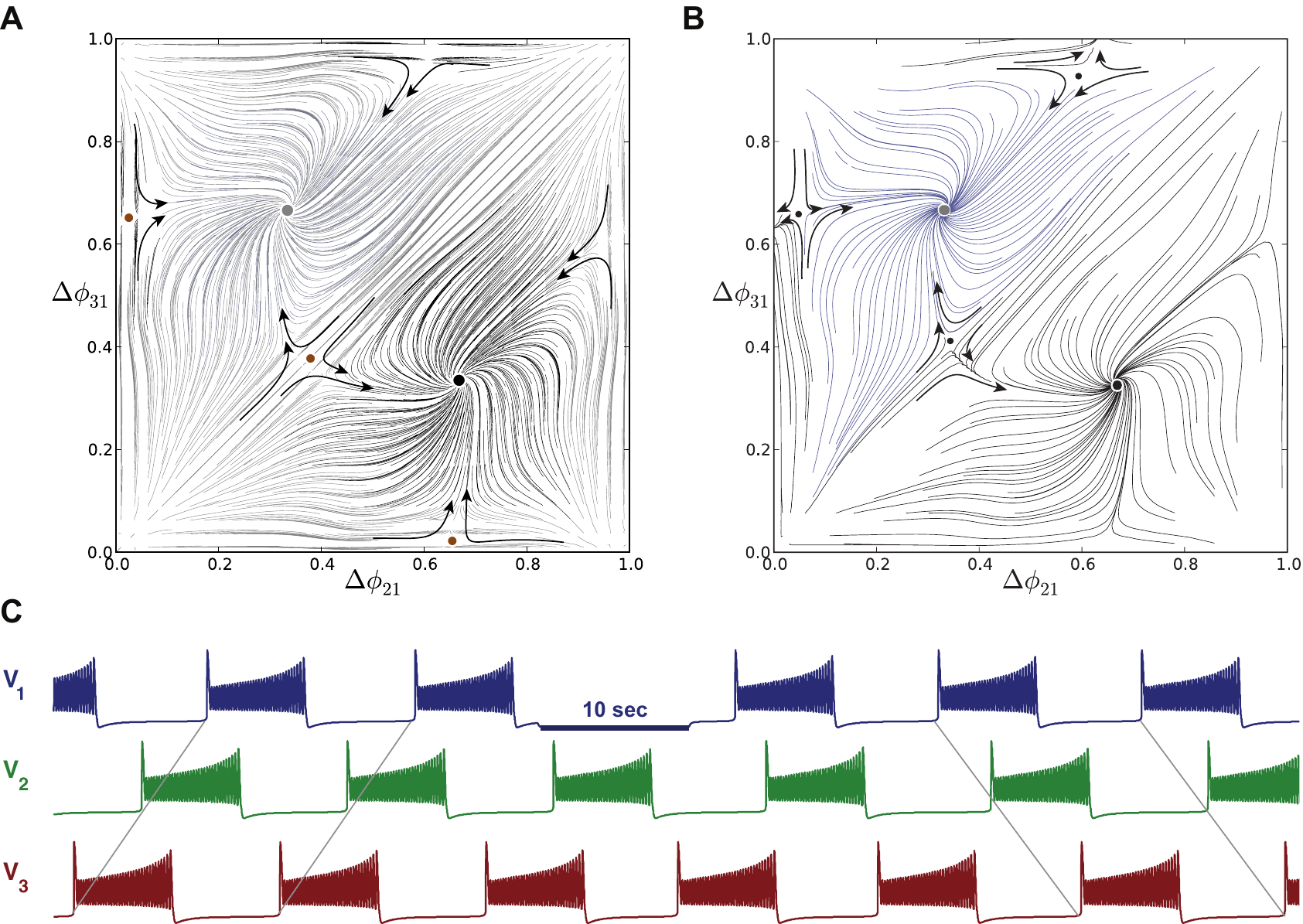}
\end{center}
\caption{ {\bf Phase lag maps in the long bursting motif and switching between two coexisting rhythms.} (A) Symmetric phase lag map for $80\%$ DC, which possesses two stable FPs  $\left (\Delta \phi_{21}, \Delta\phi_{31} \right ) = \left (\frac{2}{3},\frac{1}{3} \right )$ and $\left (\frac{1}{3},\frac{2}{3} \right)$ of equal  basins that correspond to a counter-clockwise $\OneThreeTwo$ and clockwise $\OneTwoThree$  traveling waves. The other three FPs have rather narrow basins, thus the traveling waves dominate the behavioral repertoire of the network. (B) Map corresponding to the clockwise biased  motif with  $\gas= 0.1$ reveals the asymmetric basins of the robust rhythms after three saddles have moved closer to the stable FP at $\left (\frac{1}{3},\frac{2}{3} \right)$.  (C) Bistability: switching from the counter-clockwise, $\OneThreeTwo$, to the clockwise, $\OneTwoThree$, traveling wave in this motif, after releasing  the target blue cell from hyperpolarized silence due to an external inhibitory pulse.}
\label{fig8}
\end{figure}

Two FPs around  $ \left (\Delta \phi_{21}, \, \Delta \phi_{31} \right ) =\left (\frac{1}{3},\frac{2}{3}  \right)$ and $\left (\frac{2}{3},\frac{1}{3} \right)$ correspond to clockwise and counter-clockwise traveling waves (respectively) that we denote $\OneTwoThree$  and $\OneThreeTwo$.  Here, the period of either traveling wave is broken into three episodes in which each cell is actively bursting one at a time. For example, in Fig.~\ref{fig4} for the clockwise bursting, $\OneTwoThree$, the cell ordering is~1-2-3 before the pattern repeats. The co-existence of these two waves originates from the rotational symmetry of the homogeneous motif. However, both such traveling bursting waves are not robust and therefore cannot  be observed in the motif with a short duty cycle because the corresponding FPs are repelling, so that a small perturbation will cause the phase lags of such a traveling rhythm to transition  to those corresponding to one of three ``pacemaker'' states, as shown in Fig.~\ref{fig7}.

%%%%

\subsubsection*{Long length motif}

Next, we consider the bursting motif that has a longer duty cycle of $80\%$. This duty cycle is achieved by setting $\vks=-0.0225\mathrm{V}$, which brings the cells closer to the boundary separating bursting and tonic spiking activities (Fig.~\ref{fig3}). The corresponding return map for the phase lags is shown in Fig.~\ref{fig8}A. There are two equally dominating stable FPs, $\left (\Delta \phi_{21}, \Delta\phi_{31} \right ) \approx \left (\frac{2}{3},\frac{1}{3} \right)$ and $\left (\frac{1}{3},\frac{2}{3} \right)$, corresponding to the now highly robust counter-clockwise $\OneThreeTwo$ and clockwise $\OneTwoThree$  traveling waves.

Figure~\ref{fig8}B illustrates the waveforms, as well as the bistability of the motif initially producing the counter-clockwise, $\OneThreeTwo$, traveling wave that reverses into the clockwise one, $\OneTwoThree$, after a 10 second inhibitory pulse ended and released the blue reference cell to initiate a burst.

%%%

\subsubsection*{Medium length motif}

To complete the examination of the influence of duty cycle on the repertoire and robustness of bursting outcomes of the homogeneous motif, we now consider the case of the medium-length duty cycle, $50\%$, at $\vks=-0.021\mathrm{V}$ (the middle interval shown in Fig.~\ref{fig3}).

\begin{figure}[h!]
\begin{center}
\includegraphics[width=0.99  \textwidth]{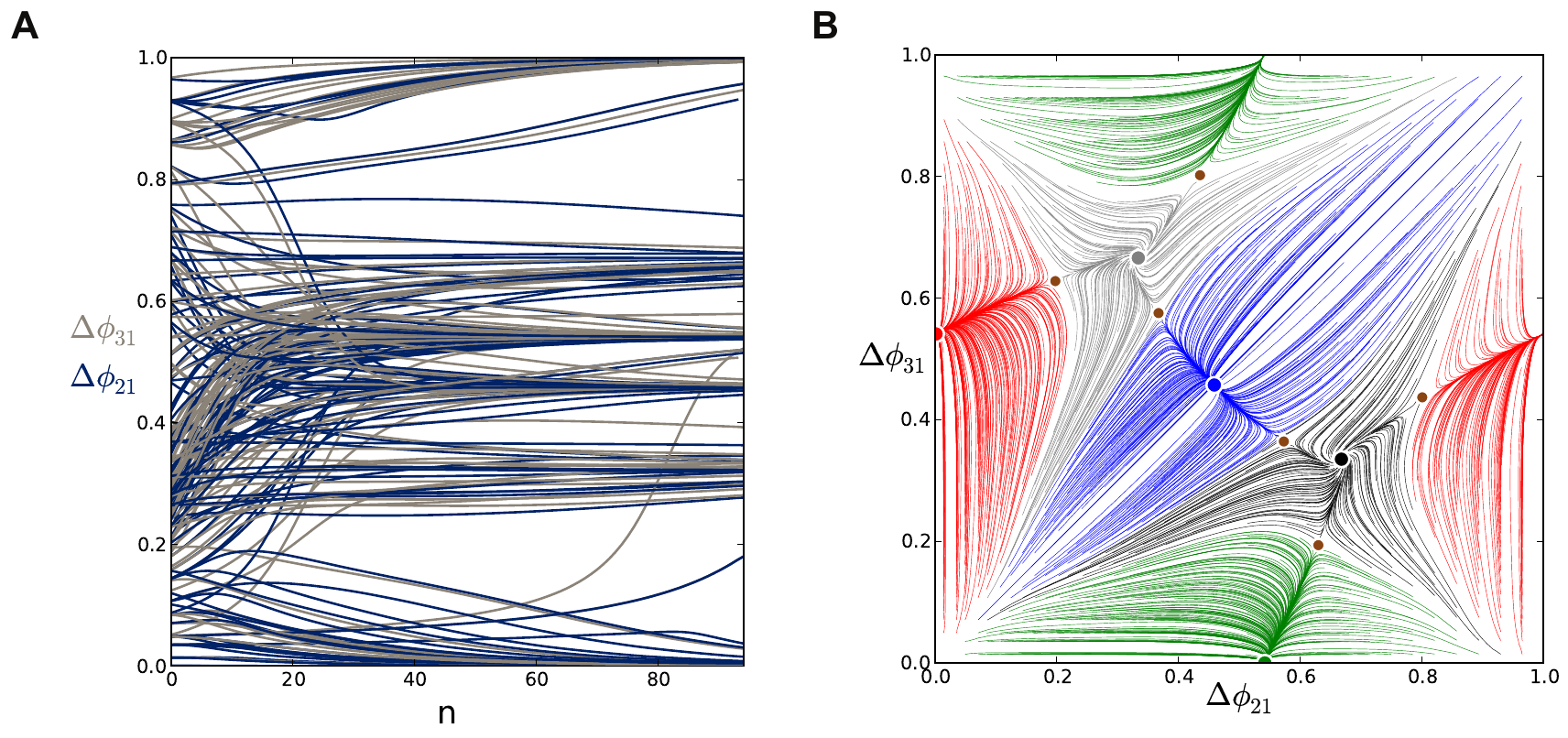}
\end{center}
\caption{ {\bf Time evolutions of voltage traces and phase lag map for the medium bursting motif.} (A) Transients of the phase lags, $\Delta \phi_{31}$ (gray) and $\Delta \phi_{21}$ (blue), converging to several
phase locked states after 90 burst cycles in the medium bursting motif with $50\%$ DC with $\gsyn=0.0005$.
(B)  The phase lag Poincar\'e map revealing five stable FPs: red dot at $(0, \, \frac{1}{2})$, green $(\frac{1}{2}, \, 0)$, blue ($\frac{1}{2},\frac{1}{2}$), black $(\frac{2}{3},\frac{1}{3})$ and purple
($\frac{1}{3},\frac{2}{3}$), corresponding to the anti-phase $\ThreePerpOneTwo$, $\TwoPerpOneThree$, $\OnePerpTwoThree$ bursts, and traveling
clockwise $\OneTwoThree$  and counter-clockwise  $\OneThreeTwo$ waves; the attractor basins of the same colors are subdivided by separatrices of six saddles (smaller brown dots). }
\label{fig9}
\end{figure}

Similarly to Figure~\ref{fig5}A, Figure~\ref{fig9}A illustrates the evolution of $\Delta \phi_{21}$ and $\Delta \phi_{31}$ (shown in blue and gray colors) from initial conditions uniformly distributed over the unit interval. One can observe transients ultimately converging to multiple constant  phase locked states. The corresponding map $\Pi$ is presented in Fig.~\ref{fig9}B. In contrast to the case of short and long bursting motifs, the map for the medium bursting motif with weak homogeneous connections reveals the coexistence of five stable FPs: the red one at $\left (0, \,\frac{1}{2}\right )$, the green one at $\left (\frac{1}{2}, \, 0 \right )$, the blue one at $\left (\frac{1}{2},\frac{1}{2} \right )$, the black one at $\left (\frac{2}{3},\frac{1}{3} \right)$ and the gray one at $\left ( \frac{1}{3},\frac{2}{3}\right)$. These FPs represent,  correspondingly,  five robust polyrhythms:
the anti-phase $\ThreePerpOneTwo$, $\TwoPerpOneThree$, $\OnePerpTwoThree$ bursting patterns, and two traveling waves,
clockwise, $\OneTwoThree$,  and counter-clockwise, $\OneThreeTwo$.

%%%%%%%% FIGURE 9 ABOUT HERE
%\vspace{0.2cm}{\it FIGURE 9 around here} \vspace{0.2cm}

%%%%%%% FIGURE 10 ABOUT HERE
%\vspace{0.2cm}{\it FIGURE 10 around here} \vspace{0.2cm}
\begin{figure}[h!]
\begin{center}
\includegraphics[width=0.99  \textwidth]{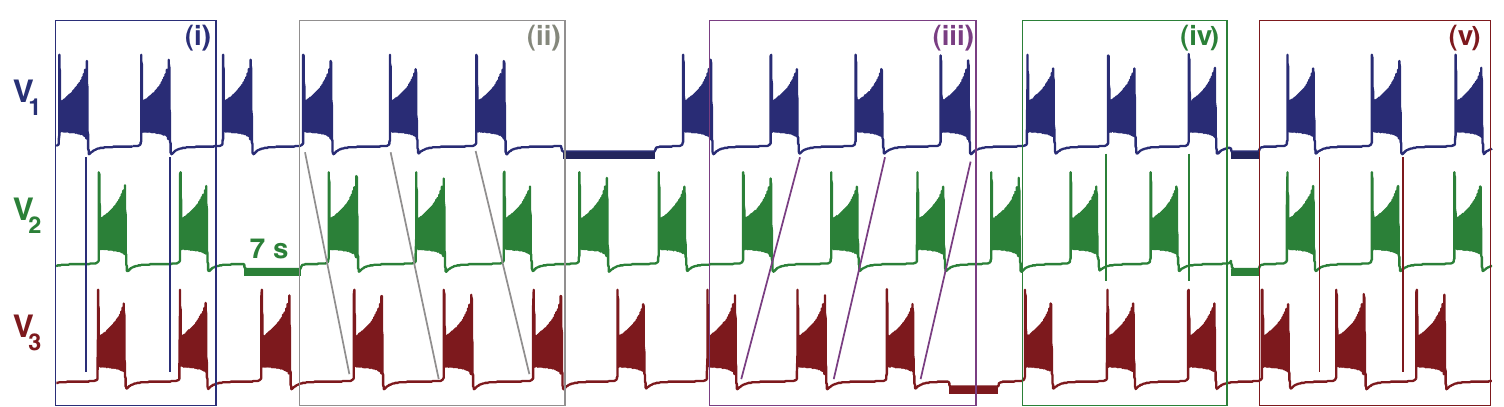}
\end{center}
\caption{ {\bf Voltage traces showing the five bursting polyrhythms in the medium duty cycle motif.} Here, we choose $g_\mathrm{syn}=5\times 10^{-3}$ to ensure short transients for the purpose of illustration. Inhibitory pulses (horizontal bars) suppress then release the targeted cells, thus causing switching between the co-existing rhythms: $\OnePerpTwoThree$ in episode (i), traveling waves $\OneTwoThree$ in (ii) and $\OneThreeTwo$ in (iii), followed by $\TwoPerpOneThree$ led by cell~2 in (iv). Having released cells~1 and 2 simultaneously, this makes cell~3 lead the motif in the $\ThreePerpOneTwo$ rhythm in the fifth episode, (v).}
\label{fig10}
\end{figure}

%%%%%%%%%%%%%%%%%%%%%

\subsection*{Asymmetric inhibitory motifs}\label{inhib_asymm}

In this section, we elucidate how and what intrinsic properties of the individual bursting cells affect the multistability of the 3-cell inhibitory motif. The answer involves an interplay between the competitive dynamical properties of individual neurons and the cooperative properties of the network. More
specifically, it relies on how close an isolated cell is to the boundary between bursting and hyperpolarized quiescence and how sensitive the post-synaptic cell is to the (even weakly) inhibitory current generated by the pre-synaptic cells. We investigate these ideas by introducing asymmetries into the coupling of our homogeneous network motif. We focus on several representative cases of asymmetrically coupled motifs with one or more altered synaptic strengths, and we will elaborate on their bifurcations as we vary the asymmetry.

%%%

\subsubsection*{From multistability to the $\OneThreeTwo$ pattern}

In this subsection, we analyze bifurcations occurring \emph{en route} from the homogeneous 3-cell motif to a rotationally-symmetric one, during which \emph{all} clockwise- and counter-clockwise-directed synapses are simultaneously increased and decreased, respectively. In the limiting case of a clockwise, uni-directionally coupled motif there is a single traveling wave. The question is: in what direction will the wave travel?

We use a new bifurcation parameter, $\gas$, which controls the rotational symmetry as the deviation from the nominal coupling strengths, $\gsyn=5 \times 10 ^{-4}$, such that $\gsyn (1 \pm \gas)$ and $0 \le \gas \le 1$. The limit $\gas \rightarrow 1$ corresponds to the unidirectional case.
Recall that initially, at $\gas=0$, both the traveling waves $\OneTwoThree$ and  $\OneThreeTwo$ are unstable in the short bursting motif with  $20\%$ DC. Then, the network can only generate the $\OnePerpTwoThree$, $\TwoPerpOneThree$, $\ThreePerpOneTwo$ pacemaker rhythms.

\begin{figure}[h!]
\begin{center}
\includegraphics[width=0.99  \textwidth]{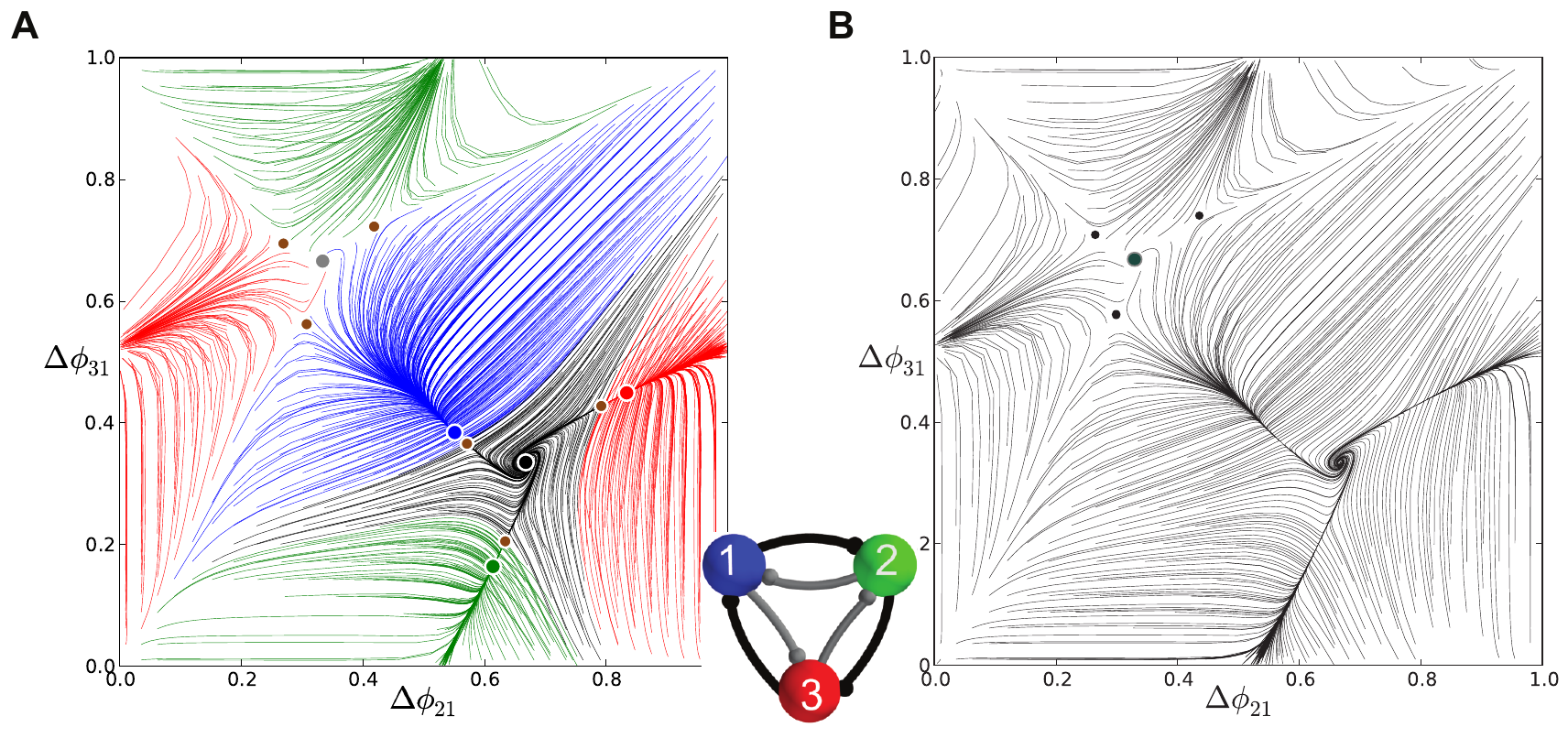}
\end{center}
\caption{ {\bf Phase lag maps near a saddle-node bifurcation for an asymmetric motif.} (A) Phase lag map for the short bursting motif and coupling asymmetry $\gas = 0.41$: the three saddles surrounding the stable
FP $(\Delta \phi_{21}, \Delta\phi_{31}) = \left (\frac{2}{3},\frac{1}{3} \right)$, are about to merge and vanish with other
three stable FPs through simultaneous saddle-node bifurcations; the FP at $\left (\Delta \phi_{21},\Delta \phi_{31} \right) =\left (\frac{1}{3},\frac{2}{3} \right)$ remains unstable. (B) For $\gas > 0.42$ the FP $\left (\Delta \phi_{21}, \Delta\phi_{31} \right) = \left (\frac{2}{3},\frac{1}{3} \right)$ becomes the only attractor of the map, which corresponds to the only robust $\OneThreeTwo$ traveling wave. The network motif is inset, where darker connections are stronger.}
\label{fig11}
\end{figure}

Figure~\ref{fig11}A depicts $\Pi$ at a critical value of $\gas=0.41$, and reveals that the FP $(\Delta \phi_{21}, \, \Delta \phi_{31} ) =\left (\frac{2}{3},\frac{1}{3} \right)$ is stable. Thus the counter-clockwise traveling  wave, $\OneThreeTwo$, is now observable
in the asymmetric motif. The value $\gas=0.41$ is a bifurcation value because further increase make the three saddles and the three initially stable FPs (blue, green and red), merge in pairs and annihilate though three simultaneous saddle-node bifurcations. After that, the FP around $\left (\frac{2}{3},\frac{1}{3}\right )$ becomes the \emph{global attractor} of the network (see Fig.~\ref{fig11}C) at $\gas=0.42$ , which produces  the single counter-clockwise $\OneThreeTwo$ traveling wave, while the FP at $\left (\frac{1}{3},\frac{2}{3} \right)$ remains unstable.

%%%%%%%%%%%% FIGURE 11 ABOUT HERE
%\vspace{0.2cm}{\it FIGURE 11 around here} \vspace{0.2cm}

\begin{figure}[h!]
\begin{center}
\includegraphics[width=0.99  \textwidth]{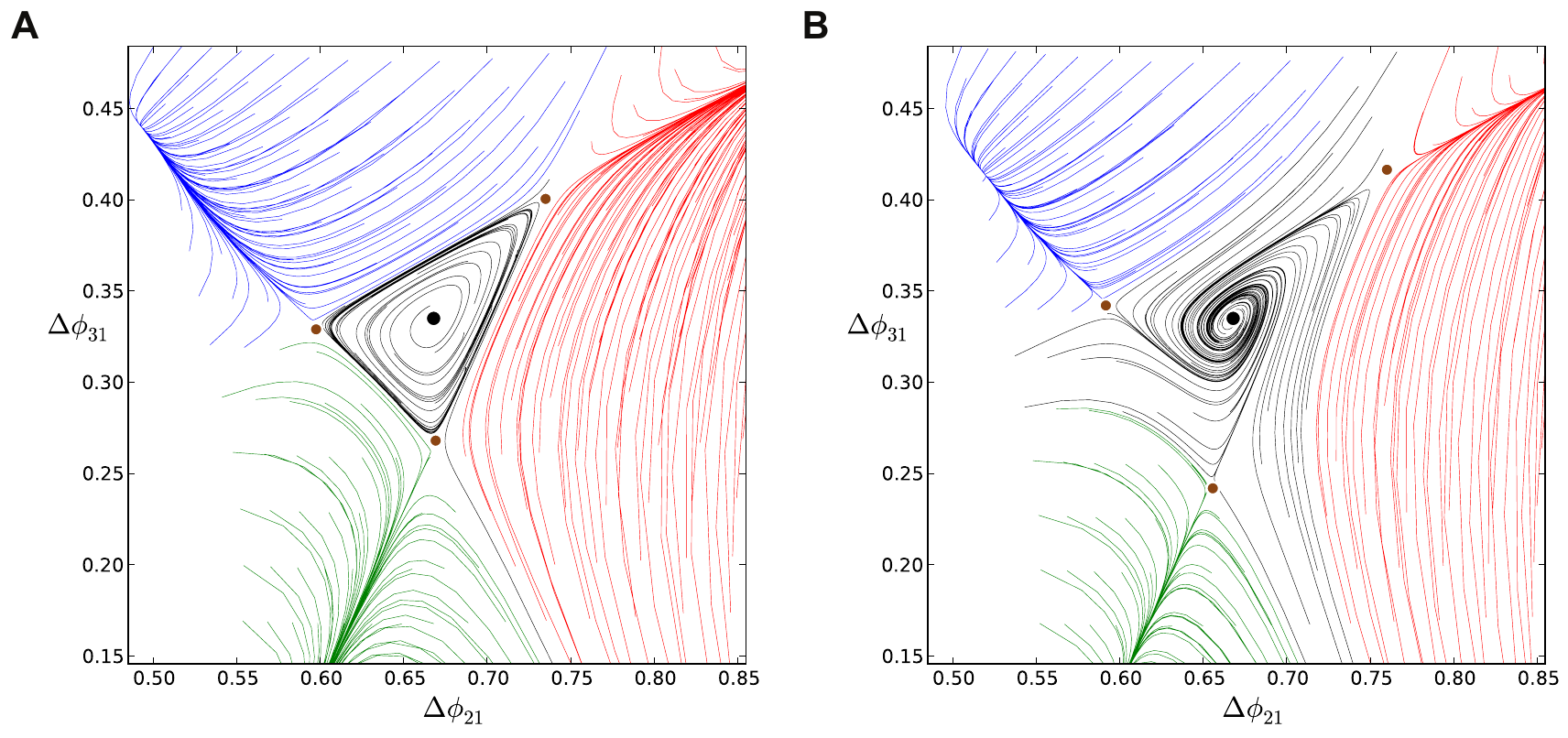}
\end{center}
\caption{ {\bf Enlargement of the phase lag map for the short bursting motifs.} (A) Case $\gas=0.185$ depicts a stable invariant circle
near a heteroclinic connection between the surrounding saddles that produces a small-amplitude phase \emph{jitter} in the voltage traces.
(B) Case $\gas=0.32$ illustrates the change in stability for the FP at
$(\Delta \phi_{21}, \Delta\phi_{31}) = \left (\frac{2}{3},\frac{1}{3} \right)$ at large values of $\gas$.}
\label{fig12}
\end{figure}

Next we have to characterize the missing stages for the transformation of the initially unstable FP, $\OneThreeTwo$, into the stable one at $\gas=0.42$. Two additional maps, shown in Fig.~\ref{fig12}, focus on the area near this point, and shed light onto the intermediates in the bifurcation sequence.
Figure~\ref{fig12}A depicts an enlargement of $\Pi$ at $\gas=0.185$. It shows a stable invariant curve near
a heteroclinic connection involving all three saddles around the FP $\left (\frac{2}{3},\frac{1}{3} \right )$.
In this figure, the FP near the center of the plot is still unstable. This indicates that the invariant curve
has emerged from the heteroclinic connection at a smaller value of the parameter $\gas$.  The stable invariant curve is associated with the appearance of slow phase ``jitters'' demonstrated by the $\OneThreeTwo$ rhythm in voltage traces.

As $\gas$ is increased further, the stable invariant curve shrinks down and collapses into the unstable FP
$\left (\frac{2}{3},\frac{1}{3} \right)$ making it stable through a secondary supercritical Andronov-Hopf (otherwise known as a torus bifurcation) as shown in Fig.~\ref{fig12}B.

%%%%%%%%%%%%% FIGURE 12 ABOUT HERE
%\vspace{0.2cm}{\it FIGURE 12 around here} \vspace{0.2cm}

%%%%%%%%%%

\subsubsection*{Bifurcations in the motif with one asymmetric connection}

%%%%% FIGURE 13 ABOUT HERE 
%\vspace{0.2cm}{\it FIGURE 13 around here} \vspace{0.2cm}
\begin{figure}[h!]
\begin{center}
\includegraphics[width=0.99  \textwidth]{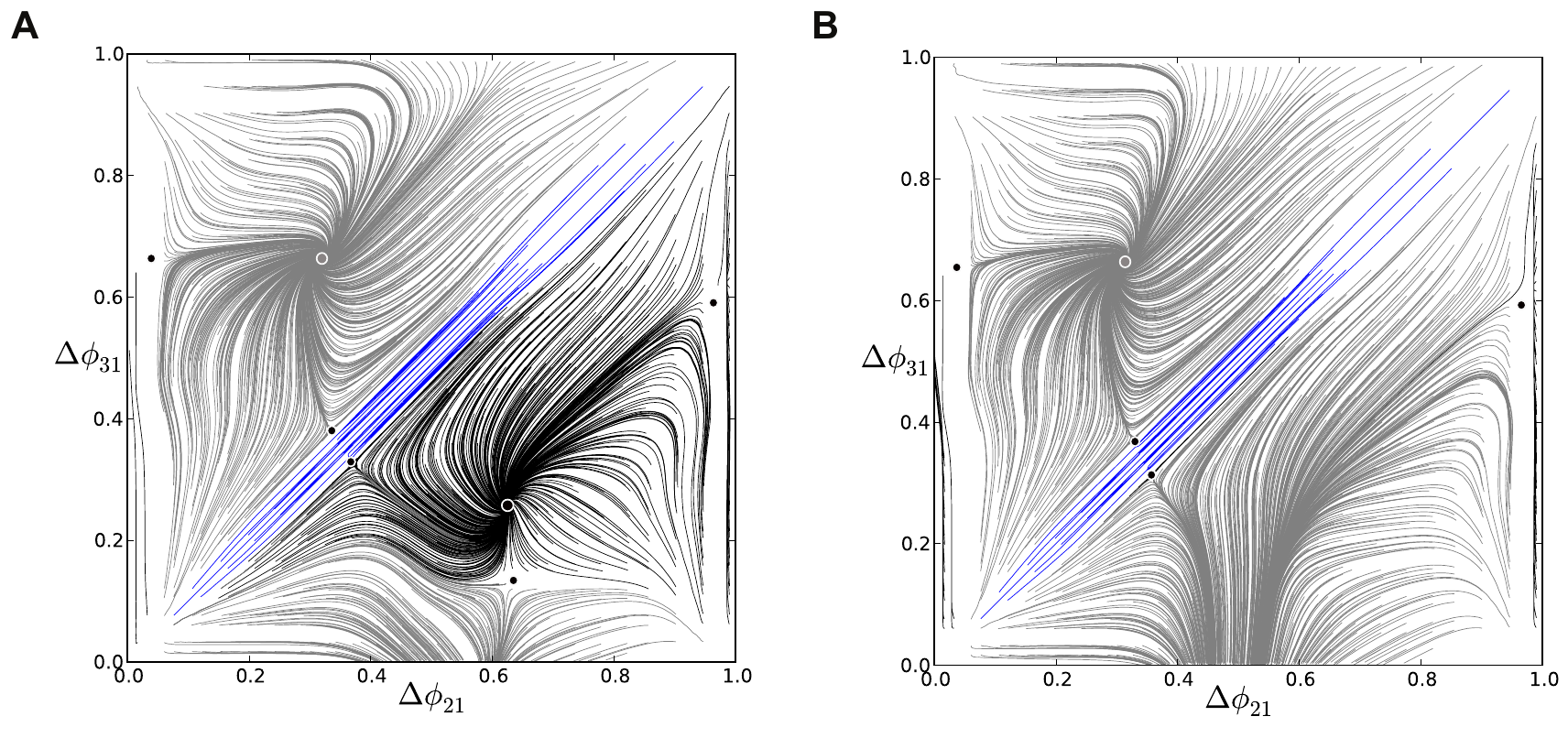}
\end{center}
\caption{ {\bf Phase lag maps for the long bursting motif with single connection asymmetry.} (A) The map for case $\gsyn_{31}=1.5 \gsyn$ possesses two attractors: one dominant at $\left (\frac{1}{3},\frac{2}{3} \right)$, and another at $\left (\frac{2}{3},\frac{1}{3} \right)$ with a smaller basin; note a saddle point in the proximity of the latter, which is a precursor of a saddle-node bifurcation. (B) Case $\gsyn_{31}=2 \gsyn$, which has a single attractor corresponding to the clockwise $\OneTwoThree$  traveling wave.}
\label{fig13}
\end{figure}

%%%%% FIG 14 around here
%\vspace{0.2cm}{\it FIGURE 14 around here} \vspace{0.2cm}
\begin{figure}[h!]
\begin{center}
\includegraphics[width=0.99  \textwidth]{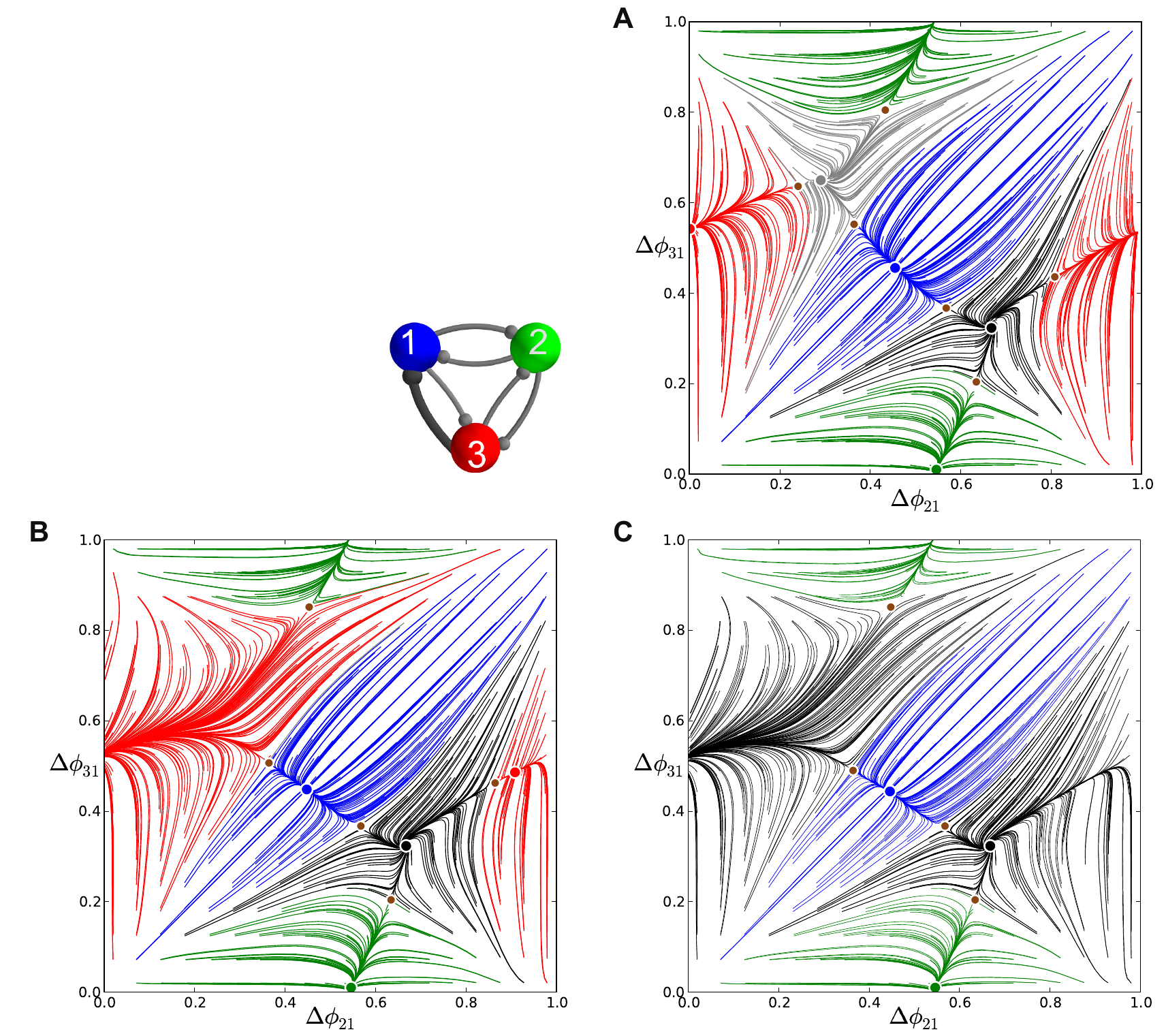}
\end{center}
\caption{ {\bf Transformation stages of the phase lag maps for an asymmetric medium bursting motif.} For the network motif shown (darker connections are stronger), a single connection 
$\gsyn_{31}$ increases from  $1.04 \gsyn$ in (A), to $1.4 \gsyn$ in (B), to $1.6 \gsyn$ in (C). In (A), the saddle between the FPs $\left (\Delta \phi_{21}, \Delta\phi_{31} \right) = \left(0,\frac{1}{2} \right )$ and $\left(\frac{1}{3},\frac{2}{3} \right)$
moves closer to the latter, then annihilates through a saddle-node bifurcation. In doing so, the attractor basin of the dominant red FP at $(0,\,\frac{1}{2})$ widens after absorbing the basin of the vanished FP in (B). In (C) a second saddle-node bifurcation annihilates the red FP.  While the counter-clockwise and reciprocal connections between  cell~2 and cells~1 and 3 remain intact, the other three stable FPs,
blue at $\left (\frac{1}{2},\,\frac{1}{2} \right )$, green $\left (\frac{1}{2},\,0 \right)$ and $\left(\frac{2}{3},\frac{1}{3} \right)$, persist in the map.}
\label{fig14}
\end{figure}

The homogeneous 3-cell motif has six independent connections, due to permutation properties we can limit our consideration of asymmetrically coupled motifs only to a few principle cases  without loss of generality. First under consideration is the motif with a single synaptic connection, $\gsyn_{31}$, from cell~3 to cell~1, being made stronger. 

 We first consider a perturbation to the homogeneous motif comprised of long bursting cells  where just a single uni-directional connection, for instance from cell 2 to 3, is strengthened. To do this, we increase the coupling stenght $\gsyn_{31}$ from the the nominal value, $5 \times 10^{-4}$, through $1.5\gsyn$, to  $2 \gsyn$. This is effectively equivalent to increasing the parameter $\vks$ only for cell~3, thus pushing it toward the quiescence boundary and extending its interburst intervals. The corresponding maps are shown in Fig.~\ref{fig13}. We observe that the initial increase of  $\gsyn_{31}$ breaks the clockwise symmetry of the motif and makes the stable node at $\left (\frac{2}{3},\frac{1}{3} \right)$ and a saddle come together. This motion further shrinks the attractor basin of the $\OneThreeTwo$ pattern. When $\gsyn_{31}$ is increased to $2\gsyn$, both FPs have annihilated through a saddle-node bifurcation. In the aftermath, the unperturbed FP at  $\left (\frac{1}{3},\frac{2}{3} \right)$ remains the unique attractor of such an map. In turn, the asymmetric motif can stably produce the single bursting pattern, which is the $\OneTwoThree$  traveling wave.

 As our case study throughout the rest of the paper, we use the non-homogeneous 3-cell motifs composed of bursting cells with $50\%$ duty cycle at $\vks=-0.021\mathrm{V}$.  Figure~\ref{fig14} depicts the stages of transformation of the phase lag maps for the motif with the connection $\gsyn_{31}$ increasing from  $1.04 \gsyn$ and $1.4 \gsyn$ through $1.6 \gsyn$. Inset~A  of Fig~\ref{fig14} shows how the variations in $\gsyn_{31}$ first break the clockwise rotational symmetry that underlies the existence of the corresponding traveling wave. As $\gsyn_{31}$ is increased to $1.04 \gsyn$ the saddle between the FPs $\left (\Delta \phi_{21}, \Delta\phi_{31}  \right) = \left (0,\frac{1}{2} \right)$ and $\left (\frac{1}{3},\frac{2}{3} \right)$ shifts closer to the one corresponding to the $\OneTwoThree$ wave. 
A further increase of  $\gsyn_{31}$ makes the saddle and the stable FP at $(\frac{1}{3},\frac{2}{3})$ annihilate through a saddle-node bifurcation. This widens the attractor basin (colored red in the figure) of the most robust FP at $\left (0,\,\frac{1}{2} \right )$  after the clockwise traveling wave has been eliminated at $\gsyn_{31}=1.4 \gsyn$, as shown in Fig.~\ref{fig14}B.  At this value of  $\gsyn_{31}$, the $\ThreePerpOneTwo$ rhythm dominates over the remaining bursting rhythms because the red cell~3 produces more inhibition than the other two. To justify this assertion we point out that another motif, with weakened clockwise connections ($\gsyn_{12}=\gsyn_{23}=0.9 \gsyn$) generates the identical Poincar\'e return map to the one shown in Fig.~\ref{fig14}A.

In the $\ThreePerpOneTwo$ rhythm, cell~3 bursts in anti-phase with the synchronous cells~1 and 2 that receive evenly balanced influx of inhibition from cell 3. 
This is no longer the case after the connection $\gsyn_{31}$ is made even stronger, so that the active cell~3 cannot hold both quiescent the postsynaptic cells~1 and 2 due to uneven coupling weights $\gsyn_{31}=1.6 \gsyn_{32}$ in the motif. One can see from the corresponding map Fig.~\ref{fig14}B that the red FP at $(0,\,\frac{1}{2})$ is approached by a saddle point from the left at $\gsyn_{31}=1.4 \gsyn$. The map 
in Fig.~\ref{fig14}C reveals that increasing  $\gsyn_{31}$ through $1.6 \gsyn$ causes a drastic change in the motif: the dominant red FP  has vanished through a subsequent saddle-node bifurcation and so has the $\ThreePerpOneTwo$ rhythm.      

With a single asymmetric connection, the structure of the phase lag map remains intact. However, the figure shows that the counter-clockwise wave has become the most robust rhythm, as the corresponding FP at $(\frac{2}{3},\frac{2}{3})$ has the largest attractor basin in the initial phase distribution.

%%%%% FIG 15 around here
%\vspace{0.2cm}{\it FIGURE 15 around here} \vspace{0.2cm}
\begin{figure}[h!]
\begin{center}
\includegraphics[width=0.99  \textwidth]{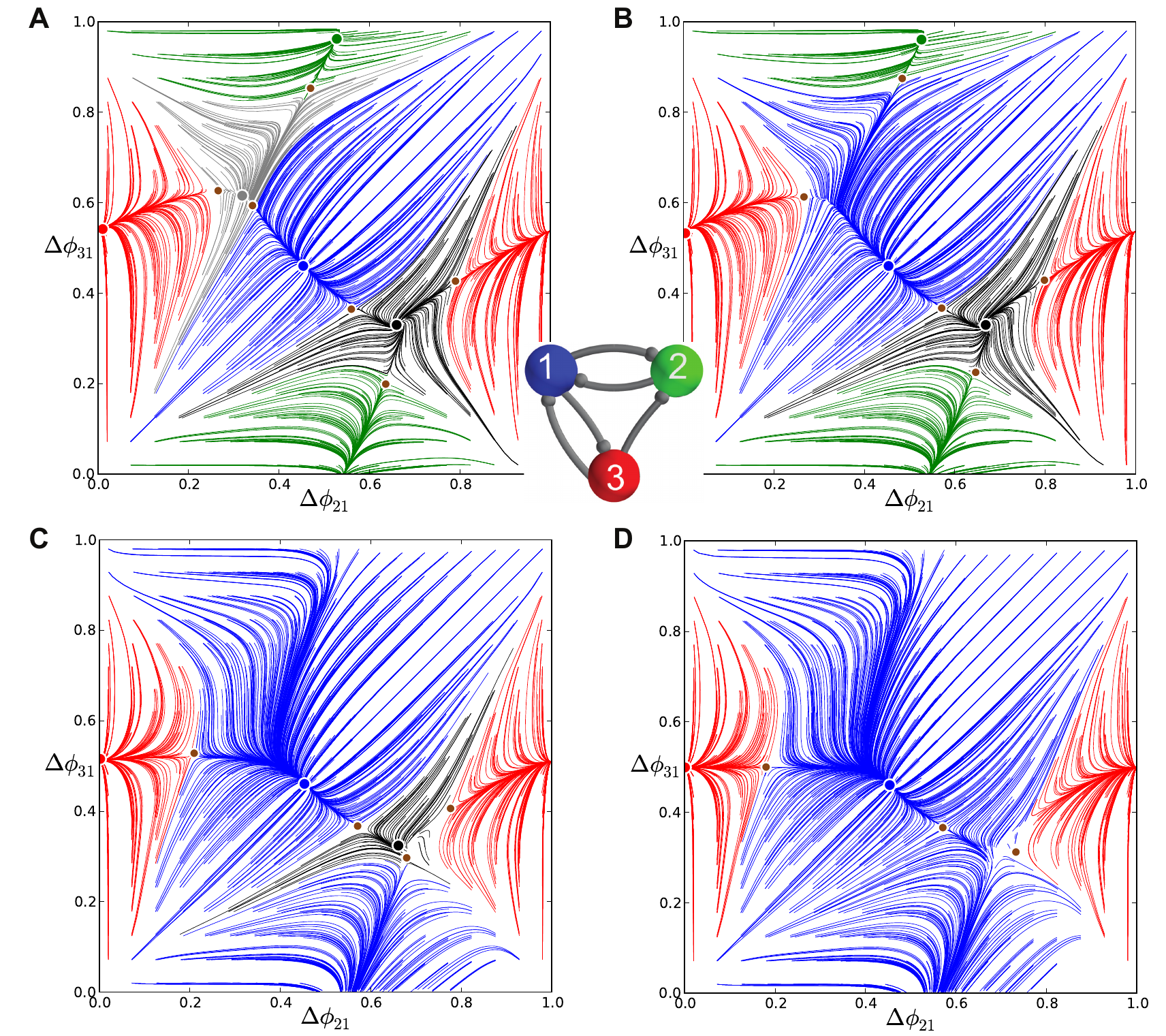}
\end{center}
\caption{ {\bf Transformation stages of the phase lag maps for the pyloric circuit motif.} Here, a single connection $\gsyn_{23}$ decreases from  $0.9 \gsyn$, $0.6$ and $0.2 \gsyn$ through to $0$ in (A)-(D), respectively. Going from (A) to (B), a triplet of saddle-node bifurcations eliminate first the clockwise $\left (\frac{1}{3},\frac{2}{3} \right )$ FP, and then subsequently 
the green FP at $\left(\frac{1}{2},\, 0 \right )$ in (B) to (C).  The growing domain of the dominant blue FP at 
$\left(\frac{1}{2},\,\frac{1}{2} \right )$ widens further from (C) to (D) after the stable counter-clockwise, $(\frac{2}{3},\frac{1}{3})$, FP is annihilated through the final saddle-node bifurcation.} 
\label{fig15}
\end{figure}

%%%%%

\subsubsection*{Pyloric circuit motif}

As an example, we examine bifurcation scenarios that occur as we transition to a heterogeneous motif that resembles the crustacean pyloric circuit with one inhibitory connection missing \cite{Marder1994752,CPG,Prinz2003,M12}. Such a network can be also treated as a sub-motif of a larger crustacean stomatogastric network \cite{CPG}.

The transformation stages are singled out in Fig.~\ref{fig15}, which shows the 
bifurcations of the FPs in the phase lag maps. As in the previous case, decreasing a single either  clockwise or counter-clockwise directional connection removes the corresponding FP at $\left ( \frac{1}{3},\frac{2}{3} \right )$ or $\left (\frac{2}{3},\frac{1}{3} \right )$, respectively. 
In this given case, it is the stable clockwise $(\frac{1}{3},\frac{2}{3})$ FP that vanishes though a saddle-node bifurcation after $\gsyn_{23}$  is decreased below $0.9 \gsyn$. Meanwhile, for $\gsyn_{23}< 0.86 \gsyn$, cell~2 cannot maintain the synchrony between cells~1 and 3 in the 
$\TwoPerpOneThree$ rhythms, which is explained by a similar argument. This assertion is supported by the phase lag maps in Fig.~\ref{fig15}B-C: one of  the saddles shifts toward to the green FP at $(\frac{1}{2},\, 0)$  and annihilates it though a subsequent saddle-node bifurcation as $\gsyn_{23}$ is decreased through $0.85 \gsyn$. The principal distinction from the prior case is that one connection, $\gsyn_{31}$,  is made twice as strong as the others in the prior case, while here we completely remove 
a single connection in the limit $\gsyn_{23}=0$. A consequence is that 
the basin of the stable FP at $\left (\frac{2}{3},\frac{1}{3} \right)$ breaks down after it vanished through the the third saddle node bifurcation that occur with  the single connection been taken out, even while the three counter-clockwise connections remain intact.  Its ``ghost'' remains influential, however, for some initial phase lags the motif can generate a long transient episode resembling the $\OneTwoThree$ traveling wave. This wave eventually transitions into the dominant anti-phase $\OnePerpTwoThree$ rhythm that coexists with the less robust $\ThreePerpOneTwo$ rhythm. In the phase plane, the ``ghost'' is located in a narrow region of transition between two saddle thresholds separating the attractor basins, blue and red, of the remaining stable FPs at $\left (\frac{1}{2},\,\frac{1}{2} \right)$ and $\left (0,\,\frac{1}{2}\right )$. Finally, removing the $\gsyn_{23}$-connection leaves the red attractor at $\left (0,\,\frac{1}{2} \right )$  and its basin intact in Fig.~\ref{fig15}D.

%%%%%%%%%%%%%%%%%

%%%%% FIG 16 around here
%\vspace{0.2cm}{\it FIGURE 16 around here} \vspace{0.2cm}
\begin{figure}[h!]
\begin{center}
\includegraphics[width=0.99  \textwidth]{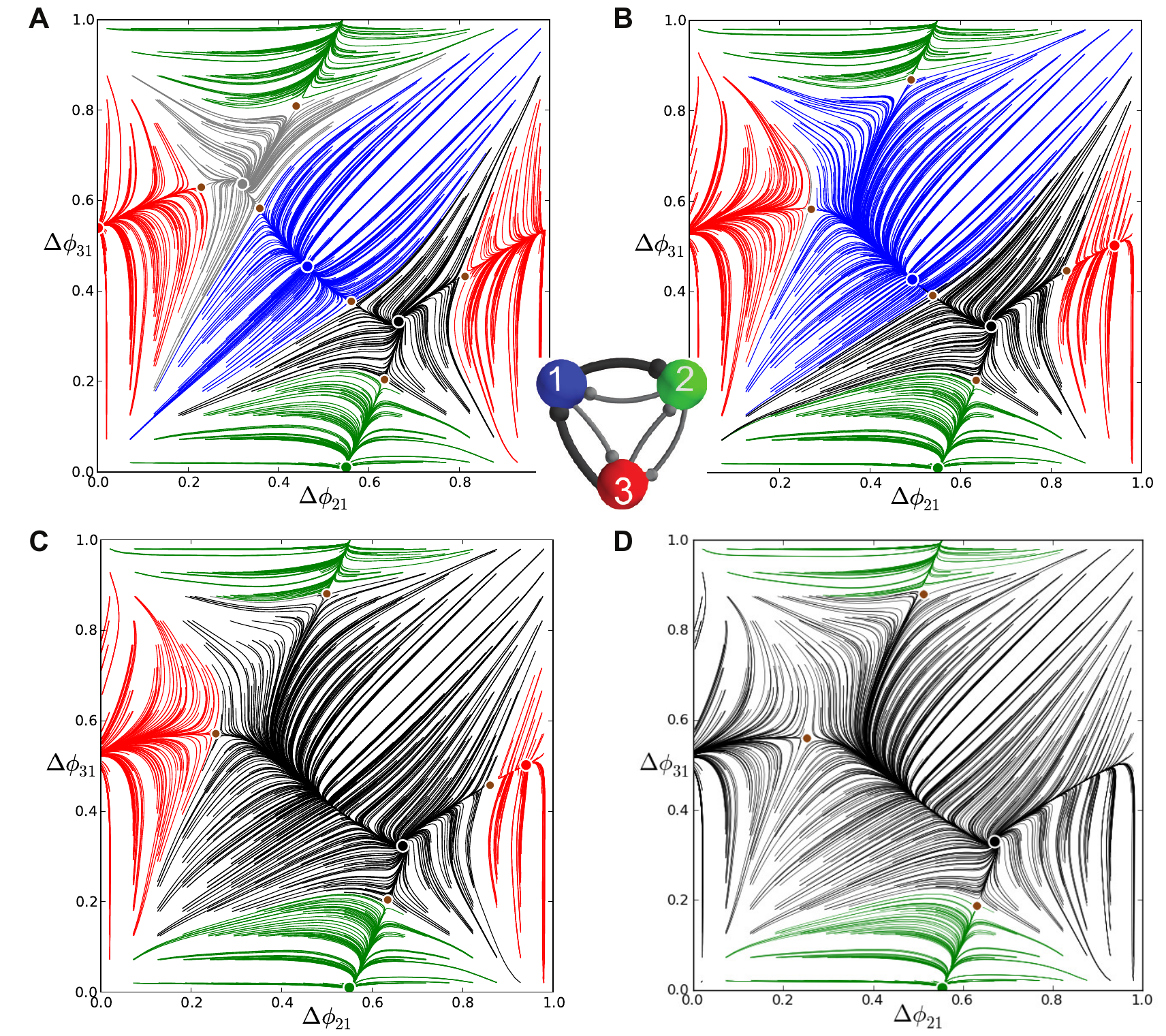}
\end{center}
\caption{ {\bf Transformation stages of the phase lag maps for a motif with uni-directional asymmetry.} Two connections $\gsyn_{31}$ and $\gsyn_{12}$ are strengthened from  $1.03 \gsyn$ through $1.5\gsyn$. 
Due to the uni-directional symmetry breaking, the map first loses the clockwise, $\left (\frac{1}{3}, \frac{2}{3} \right)$, FP (light gray) after it merges with a saddle at $1.05 \gsyn$,  then the blue $\left (\frac{1}{2},\, \frac{1}{2} \right)$ and the red $\left (\frac{1}{2},\, 0 \right)$ FPs disappear
through saddle-node bifurcations at $1.35 \gsyn$ and $1.45 \gsyn$, respectively. As the counter-clockwise connections remain the same, the presence of the remaining FPs at $\left (\frac{2}{3},\,\frac{1}{3} \right)$ and  $\left (\frac{1}{2},\, 0 \right )$ on the  torus  guarantees that the $\OneThreeTwo$ traveling wave and the $\TwoPerpOneThree$ rhythm persist in the motif's repertoire.}
\label{fig16}
\end{figure}

%%%

\subsubsection*{Two asymmetric connections: uni-directional case}

Here, we examine the motif with two uni-directional connection asymmetries, for example where $\gsyn_{12}$ and $\gsyn_{31}$ are strengthened from  the nominal value to $1.5\gsyn$. The bifurcation stages of $\Pi$ are depicted in Fig~\ref{fig16}. During the transformations, the map loses
three FPs in sequence through similar saddle-node bifurcations. Because increasing $\gsyn_{31}$ and $\gsyn_{12}$ breaks the clockwise symmetry, the corresponding FP at $\left (\frac{1}{3}, \frac{2}{3} \right)$ for the counter-clockwise wave, $\OneTwoThree$,  is annihilated first at 
around $1.05\gsyn$ after merging with a saddle.  Further strengthening both corrections annihilates the blue FP at  $\left (\frac{1}{2},\, \frac{1}{2} \right)$, followed by the red FP at $(\frac{1}{2},\, 0)$. As such, the pacemaker $\OnePerpTwoThree$ and $\ThreePerpOneTwo$  
rhythms eventually are no longer available as neither cells~1 nor 3 are able to hold  
the post-synaptic counterparts in synchrony, and also because the periods of the unevenly driven cells become too different.     

The clockwise symmetry breaking does not affect counter-clockwise connections. Thus, in the map for $1.5\gsyn$, two rhythmic patterns persist: the $\OneThreeTwo$ traveling wave with a wide attractor basin and the pacemaker $\TwoPerpOneThree$ rhythm. Their associated FPs are at $(\frac{2}{3},\,\frac{1}{3})$ and $(\frac{1}{2},\, 0)$, respectively.   
It is worth noticing that the same sequence of bifurcations will not occur in the map and the motif if only the connection $ \gsyn_{23}$ is weakened instead.

%%%%
\subsubsection*{Two asymmetric connections: Unilateral dominance case}

\begin{figure}[h!]
\begin{center}
\includegraphics[width=0.99  \textwidth]{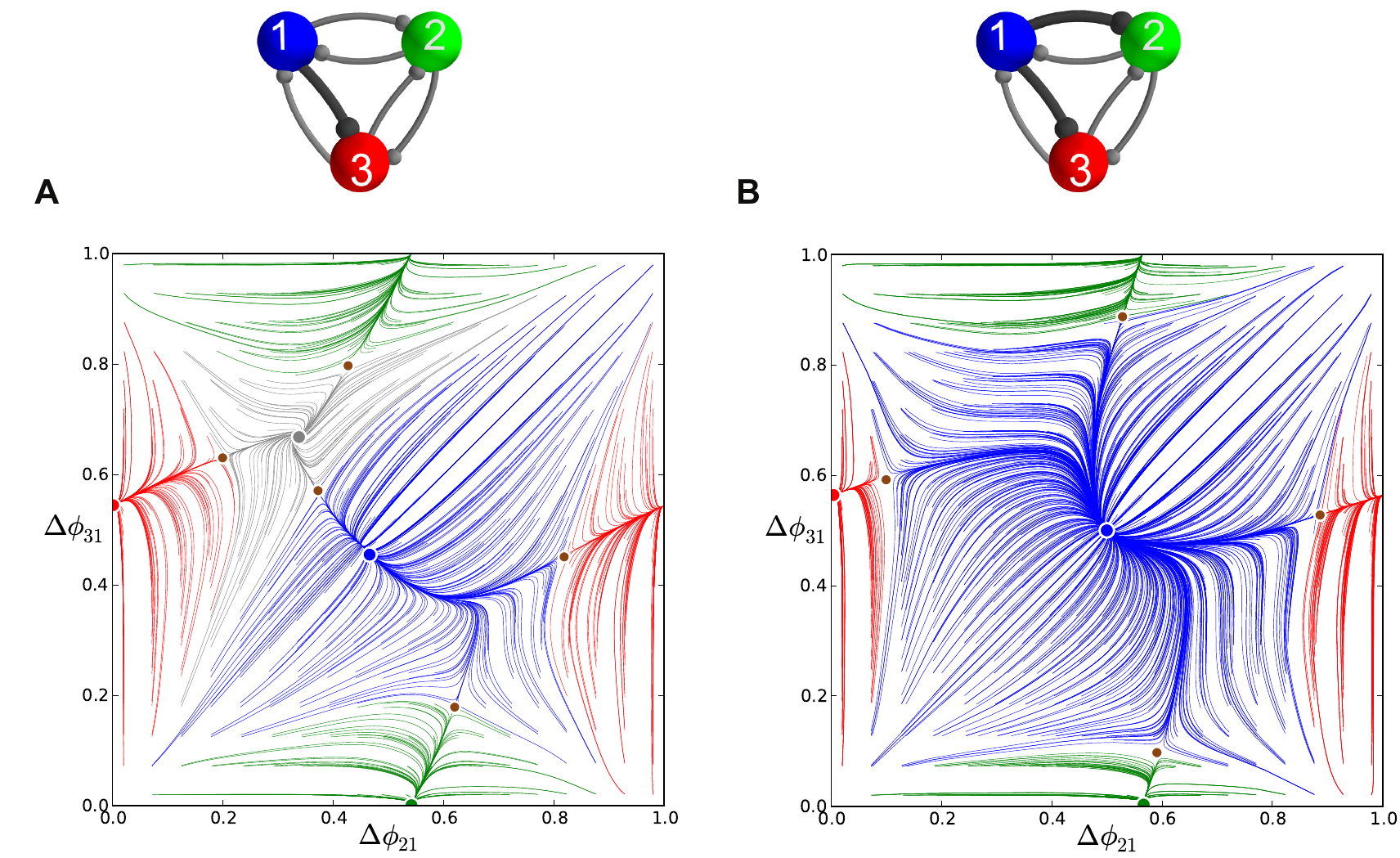}
\end{center}
\caption{ {\bf Representative phase lag maps for motifs with other connection asymmetry types, Part 1.} (A) Counter-clockwise biased motif with the single strengthened connection $\gsyn_{13}=1.1 \gsyn$ and medium duty cycle. 
The phase lag map lacks the FP at $(\frac{2}{3}, \frac{1}{3})$ and the saddle near the dominating blue FP at $(\frac{1}{2},\, \frac{1}{2})$. 
(B) Motif with a strongly inhibiting cell~1 due to two strengthened connections: $\gsyn_{12}=\gsyn_{13} = 2\gsyn$. 
The phase lag map with the strongly dominating FP at $(\frac{1}{2},\,\frac{1}{2})$ for 
the $\OnePerpTwoThree$ rhythm  whose attractor basin expands over those of the FPs corresponding to clockwise  
$\OneTwoThree$ and counter-clockwise $\OneThreeTwo$ traveling waves. This larger basin has narrowed those of the coexisting  
stable green FP at $(\frac{1}{2},\, 0)$ for the $\TwoPerpOneThree$ rhythm and the red FP at $(0,\, \frac{1}{2})$ for the $\ThreePerpOneTwo$ rhythm.}
\label{fig17}
\end{figure}

Next under consideration is a motif in which cell~1 alone produces stronger inhibitory output due to two strengthened connections, $\gsyn_{12}$ and $\gsyn_{13}$. 
Figure~\ref{fig17} depicts two snapshots of the phase spaces of the map after $\gsyn_{13}$ and then $\gsyn_{12}$ have been strengthened. 
One sees that a $10\%$ increase in inhibition   in the counter-clockwise direction breaks the rotational symmetry and therefore makes the stable FP at $\left  (\frac{2}{3},\frac{1}{3} \right)$ (corresponding to the $\OneThreeTwo$ rhythm) disappear through a saddle-node bifurcation as it merges with a saddle. As in the previous cases, the attractor basin of the stable blue pacemaker at $\left (\frac{1}{2},\,\frac{1}{2} \right)$ extends to absorb that of the former FP. As expected, since all counter-clockwise connections have remained equal in this case, the stable FP at $ \left (\frac{1}{3},\frac{2}{3} \right)$ persists, as does the  $\OneTwoThree$ traveling wave. The dominating rhythm, clockwise traveling wave $\OneTwoThree$, coexists with anti-phase $\TwoPerpOneThree$, $\ThreePerpOneTwo$ rhythms.

%%%%%% FIGURE 17 ABOUT HERE
%\vspace{0.2cm}{\it FIGURE 17 around here} \vspace{0.2cm}

Next, in addition to $\gsyn_{13}$, the second outgoing connection, $\gsyn_{12}$, from cell~1 is strengthened thus  breaking the clockwise symmetry as well. As expected, this eliminates the FP at $(\frac{1}{3},\frac{2}{3})$ and the corresponding clockwise $\OneTwoThree$ traveling pattern from the motif. Figure~\ref{fig17}B shows the map for the motif with $\gsyn_{12}=\gsyn_{13}=1.5 \gsyn$. While it retains all three ``pacemaker'' FPs, the one at $\left (\frac{1}{2},\,\frac{1}{2} \right)$ corresponding to the strongly inhibiting pre-synaptic  cell~1 possess the largest attractor basin.

We may conclude that strengthening a single directional connection, or alternatively, a simultaneous and proportional weakening coupling strengths of the two synaptic connections of the same orientation in the motif, controls one of the three
saddle points between the FP corresponding to traveling waves and the pacemaker FP corresponding to the stronger inhibiting cell.  This will eventually causes the disappearance of either point as soon as the rotational symmetry is broken after the coupling strength is increased over some critical value, which varies depending on the nominal value $\gsyn$ and the duty cycle of the bursting cells.

%%%%%%%%%%%%%%%%%%

\subsubsection*{Motifs with a stronger coupled HCO: loss of phase-locking}

%%%%%%%%% FIGURE 18 ABOUT HERE
%\vspace{0.2cm}{\it FIGURE 18 around here} \vspace{0.2cm}
\begin{figure}[h!]
\begin{center}
\includegraphics[width=0.99  \textwidth]{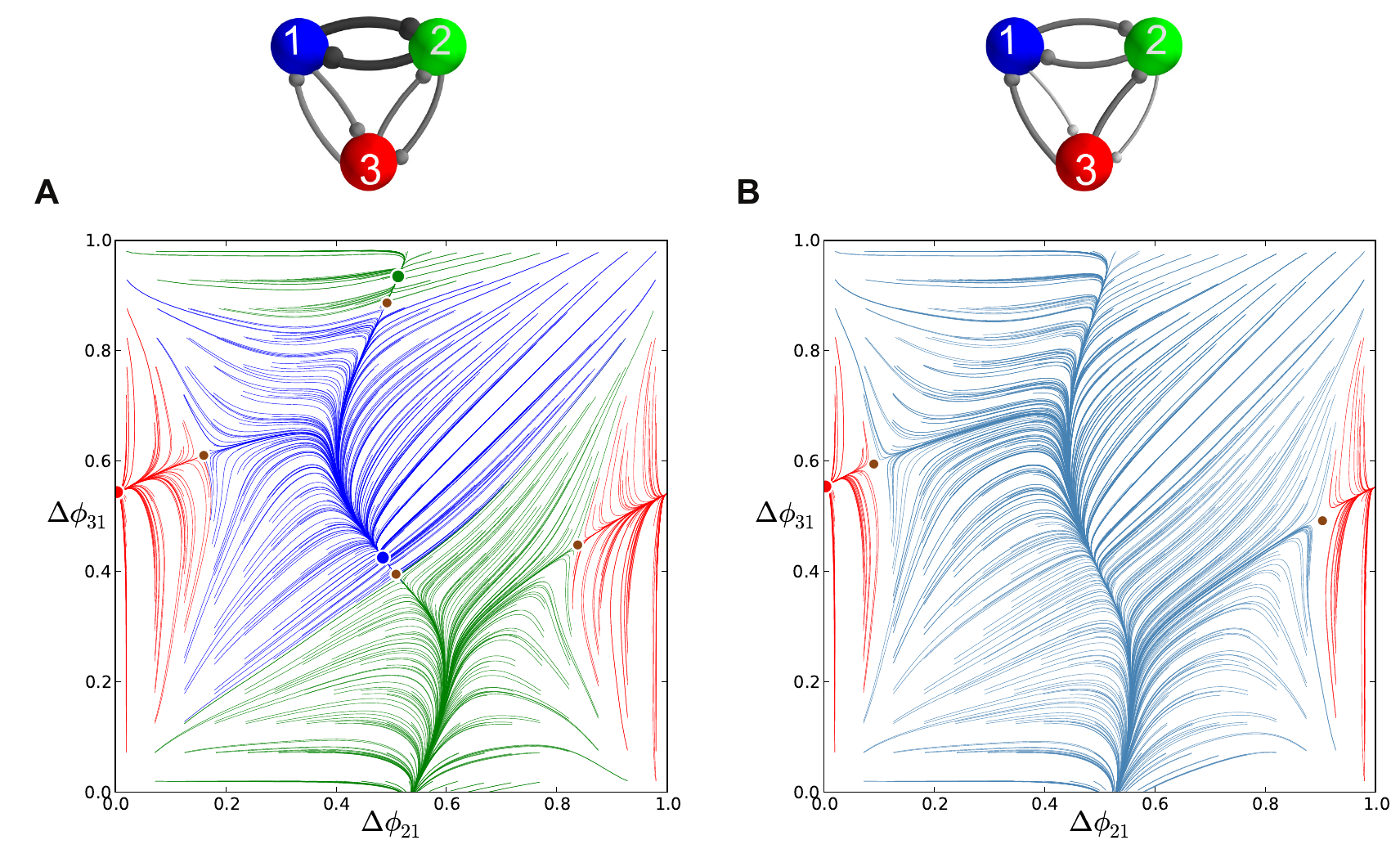}
\end{center}
\caption{ {\bf Representative phase lag maps for motifs with other connection asymmetry types, Part 2.} Motifs with two connections strengthened according to $\gsyn_{12}=\gsyn_{21}$ (A) and weakened $\gsyn_{13}=\gsyn_{23}$ (B), resulting in qualitatively similar maps. Due to the broken rotational symmetries, the maps both no longer possess FPs for the clockwise $\OneTwoThree$ and  counter-clockwise $\OneThreeTwo$ traveling waves. (C) The phase lag maps for  $\gsyn_{12}=\gsyn_{21} = 1.25\gsyn$ and for $\gsyn_{13}=\gsyn_{23} = 0.8\gsyn$.  Two large attractor basins belong to the stable (blue) FP in the middle for the $\OnePerpTwoThree$ rhythm and the stable (green) fixed point at $\left (\frac{1}{2},\, 0 \right)$ for $\TwoPerpOneThree$ rhythm. These co-exist with a smaller basin of the red fixed point at $\left (0,\,\frac{1}{2} \right)$. (D) Further increasing to $\gsyn_{12}=\gsyn_{21} = 1.5\gsyn$ in motif (A), or decreasing to $\gsyn_{13}=\gsyn_{23} = 0.6\gsyn$ in motif (B) makes the blue and green FPs vanish through consecutive saddle-node bifurcations, thus resulting in the appearance of the stable invariant curve wrapping around the torus.  The invariant circle repeatedly traverses throughout the ``ghosts'' of the four vanished FPs. Note the shrinking basin of the red FP at $\left (\frac{1}{2},\,0 \right)$  with decreasing $\gsyn_{31}=\gsyn_{32} = 0.8\gsyn$ in motif (A).}
\label{fig18}
\end{figure}

A 3-cell motif with the cells coupled reciprocally by inhibitory synapses can be viewed alternatively as a group of three half-center oscillators (HCO). Each HCO represents a pair of cells that typically burst in anti-phase, when isolated from other cells. When a HCO is symmetrically driven, even weakly, by another bursting cell, it can produce in-phase bursting, instead of out of phase bursting \cite{prl08}.

In this section, we consider transformations of rhythmic outcomes in the motif containing a single HCO with stronger reciprocally inhibitory connections,  for example,  $\gsyn_{12}=\gsyn_{21}=1.25 \gsyn$ (see  Fig.~\ref{fig18}A).  
It turns out that a 25\% increase in coupling is sufficient to break both
rotational symmetries  because it eliminates the associated FPs around $\left (\frac{1}{3}, \frac{2}{3} \right)$ and $\left (\frac{1}{3}, \frac{2}{3} \right)$ through saddle-node bifurcations.  Since both connections are strengthened simultaneously, the attractor basins of the both dominating FPs, blue near $\left (\frac{1}{2},\,\frac{1}{2} \right)$ and green near $\left (\frac{1}{2},\, 0 \right)$, widen equally. However, increasing the connections $\gsyn_{12}$ and $\gsyn_{21}$ between 
cells~1 and 2 does not affect the attractor basin of the red FP at $\left (0,\,\frac{1}{2} \right)$. In other words, the motif can still produce the co-existing $\ThreePerpOneTwo$ rhythm. 

The following bifurcation sequence involving the dominant FP differs drastically  from the saddle-node bifurcations discussed earlier. Observe from the map in Fig.~\ref{fig18}A that two saddles separating two attractor basins, have moved close to the blue and green FPs as the coupling between the HCO cells is increased to $1.5 \gsyn$.  
This is a direct indication that a further increase of the coupling strength between the strongly inhibitory cells~1 and 2 will cause two  simultaneous saddle-node bifurcations that eliminate both stable FPs.

A feature of these bifurcations of the map at the critical moment is that there are two \emph{heteroclinic connections} that bridge the saddle-node FPs on the 2D torus. The breakdown of the heteroclinic connections with the disappearance of both FPs results in the emergence of a stable invariant circle that wraps around the torus \cite{books,shilnikov2004some}. The attractor basin of the new invariant curve is bounded away from that of the red FP at $\left (0,\,\frac{1}{2} \right)$ by the stable sets (i.e., incoming separatrices) of the two remaining saddles. This motif is therefore bi-stable as the corresponding map shows two co-existing attractors.

Further increase in the coupling strength between the stronger inhibitory HCO and cell~3 cannot not qualitatively change the structure of the phase lag map, while it can have only a qualitatively effect on the size of the attractor basins of the invariant circle and the remaining FP (red).  So, weakening $\gsyn_{31}= \gsyn_{32}= 0.8\gsyn$ makes the separating saddles come closer to the red FP and hence shrink its attractor basin, as seen in Fig.~\ref{fig18}B. 

\begin{figure}[h!]
\begin{center}
\includegraphics[width=0.99  \textwidth]{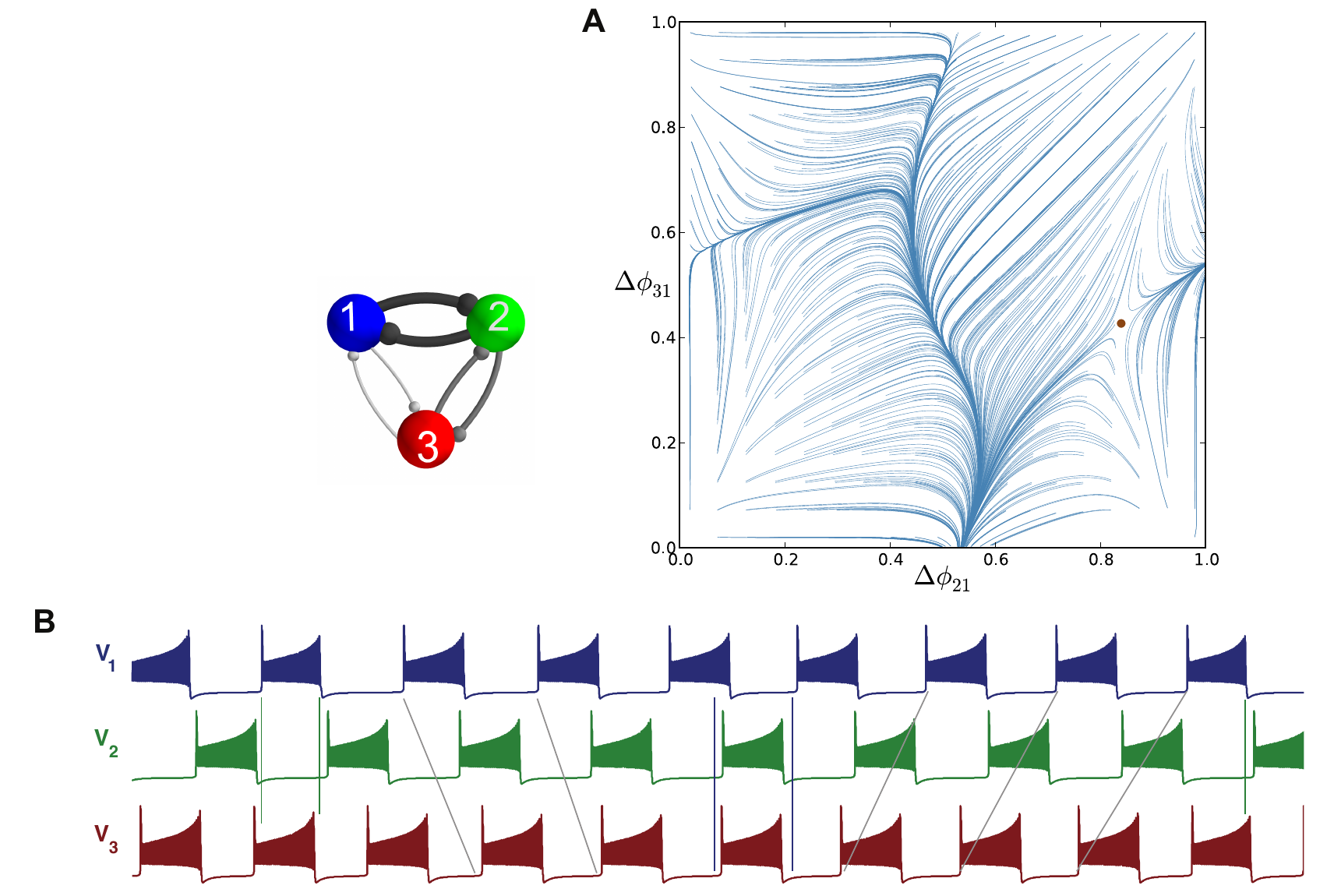}
\end{center}
\caption{ {\bf Asymmetric motifs that only exhibit phase slipping.} (A) Here, $\gsyn_{12}=\gsyn_{21} = 1.5\gsyn$ and $\gsyn_{13}=\gsyn_{31} = 0.8\gsyn$. The phase lag map possesses only one attractor:  the invariant curve corresponding to the phase slipping regime. (B) Voltage traces showing phase slipping beginning with the $\TwoPerpOneThree$ rhythm and continuously transitioning into the clockwise $\OneTwoThree$ traveling wave, followed by the $\OnePerpTwoThree$ rhythm, and being continued by the counter-clockwise $\OneThreeTwo$ traveling wave and coming back to the initial $\TwoPerpOneThree$ rhythm in nine bursting cycles.}
\label{fig19}
\end{figure}

This is not the case when either connection between cell~3 and the HCO is made sufficiently asymmetric. Depending on the connection's direction of asymmetry, such an imbalance  causes either of the two remaining saddles to come close and annihilate with the stable red FP at $\left (\frac{1}{2}, \, 0 \right)$. Figure~\ref{fig19} presents the map for this motif with weakened
reciprocal connections between cells~3 and 1:  $\gsyn_{13}=\gsyn_{31} = 0.8\gsyn$. This 
motif, comprised of three HCOs with strong, nominal and weak reciprocal connections,   no longer produces any phase-locked bursting rhythm, including $\ThreePerpOneTwo$, as the map no longer has any stable FPs. The resulting motif is monostable with a single attractor for the stable invariant curve. This curve can be characterized with a rational or irrational \emph{winding number}. The number is a rational if the invariant curve is made of a finite number of periodic points across the torus.

The occurrence of the stable invariant curve wrapping around the torus gives rise to a \emph{phase slipping} phenomenon observed in voltage traces such as those shown in Fig.~\ref{fig19}C. We define ``phase slipping'' as a repetitive rhythm with varying phase lags between the bursting cells of the motif. The period of the invariant circle depends on how far the  map with the invariant circle is from the bifurcations of ``ghost''   FPs. The ``ghosts'' make the bursting pattern with varying phase lags appear as it is composed of four sequential episodes and transitions between them.

%%%%% FIG 19 around here
%\vspace{0.2cm}{\it FIGURE 19 around here} \vspace{0.2cm}

From the top of the $\left (\Delta \phi_{21}, \Delta\phi_{31} \right)$-unit square, the curve begins with the $\TwoPerpOneThree$ rhythm continuously transitioning into the clockwise $\OneTwoThree$ traveling wave, followed by the $\OnePerpTwoThree$ rhythm, and being followed by the counter-clockwise $\OneThreeTwo$ traveling wave and finally returning to the initial $\TwoPerpOneThree$ rhythm in nine bursting cycles, which is the period of the phase slipping.  Each episode of the phase slipping rhythm can be arbitrarily large as it is controlled by the coupling strength of the specific motif connections near the corresponding saddle-node bifurcation(s). Observe that $\Delta \phi_{21} \simeq \frac{1}{2}$ on the invariant curve, i.e., while cell~1 and 2 are in anti-phase bursting, cell~3 modulates the rhythm by recurrently slowing down and advancing the HCO to generate continuously all four episodes.     

One may wonder about what determines the direction of the invariant curve on the torus and hence the order of the episodes of the shown voltage waveform. Observe that the phase slip occurs in the $\Delta \phi_{31}$ direction and that the invariant curve, unlike a FP,  has no fixed period for the whole network. Indeed, the recurrent times of this network change periodically, approximately every eight episodes. The eight episodes constituting the bursting pattern are determined by  a rational ratio of the longer HCO period (due to stronger reciprocal inhibition that extends the HCO interburst intervals) to the shorter period of pre-synaptic cell~3 (due to a weaker incoming inhibition) (see Fig.~\ref{fig19}C). 

Let us discuss another motif configuration, shown in Fig.~\ref{fig18}B, that produces  maps with stable invariant curves that wrap around the torus. These are qualitatively identical to the maps for the motif containing the strong HCO formed by cells 1 and 2 (Fig.~\ref{fig23}A). In this configuration, cell~3 receives weaker inhibition from pre-synaptic cells 1 and 2 according to  $\gsyn_{13}=\gsyn_{23} = 0.6\gsyn$. The de-stabilizing factor 0.6 turns out to be small enough to make sure that neither cell~1 nor~2 can be a pacemaker as the corresponding stable FPs have  disappeared because the period of cell~3  has become shorter than the periods of cells 1 and 2.  As the result, the map demonstrates the same stable invariant curve that ``flows'' downwards with decreasing $\Delta \phi_{31}$ phase lags.
      
\begin{figure}[h!]
\begin{center}
\includegraphics[width=0.55  \textwidth]{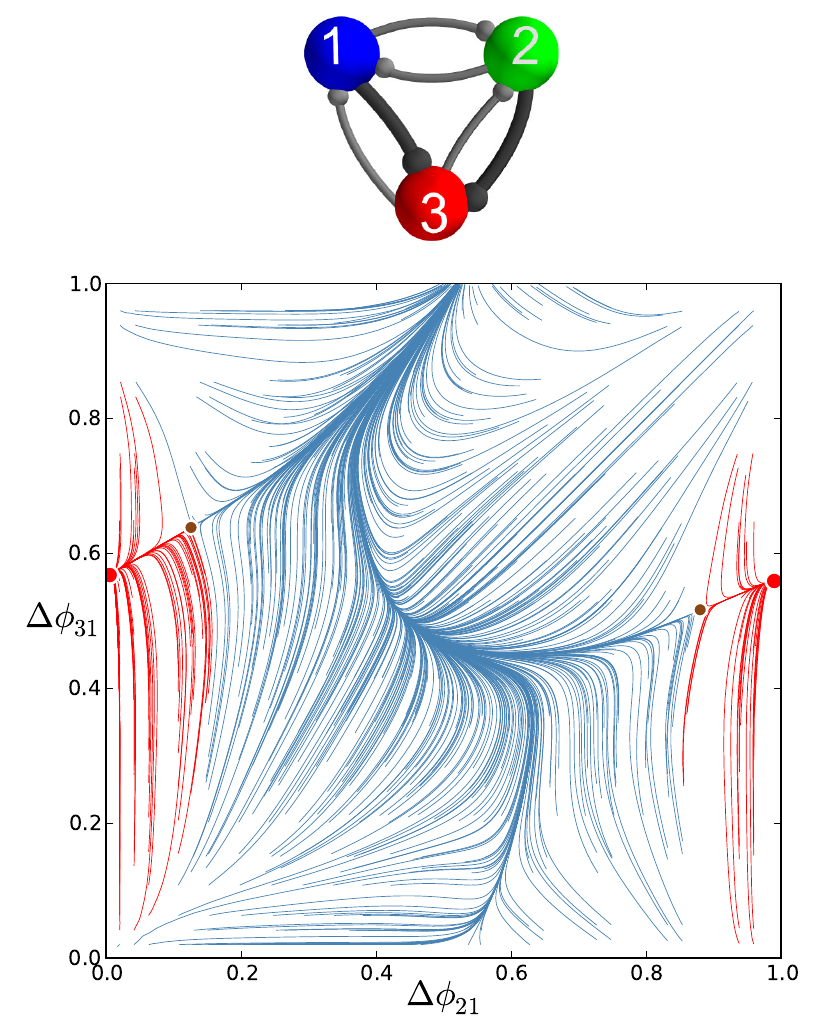}
\end{center}
\caption{ {\bf Asymmetric motif with strong connections to cell 3 shows.} (A) Motif with cell 3 receiving inhibition stronger than the nominal value: $\gsyn_{13}=\gsyn_{23} = 1.6\gsyn$. Such strong asymmetry means the map no longer possesses the traveling wave or the blue and green pacemaking FPs, similar to that shown in  Fig.~\ref{fig18}D. There is bi-stability between the two remaining attractors, i.e. the stable red FP at $\left (\frac{1}{2},\,0 \right)$ and the stable invariant curve. The stable invariant curve ``flows'' upwards, because the period of cell~3 is longer than the period of cells 1 and 2.}
\label{fig20}
\end{figure}

The direction of the stable invariant circle flowing across the 2D torus can be reversed by making cell~3 receive stronger inhibition instead of weaker inhibition relative to the other cells. An example is depicted in the phase lag map of Fig.~\ref{fig20}, where $\gsyn_{13}=\gsyn_{23} = 1.6\gsyn$.

%%%%% FIG 20 around here
%\vspace{0.2cm}{\it FIGURE 20 around here} \vspace{0.2cm}

%%%%%%%%%%%%%%
\subsubsection*{Toward control of multistability}

%%%%% FIG 21 around here
%\vspace{0.2cm}{\it FIGURE 21 around here} \vspace{0.2cm}
\begin{figure}[h!]
\begin{center}
\includegraphics[width=1.  \textwidth]{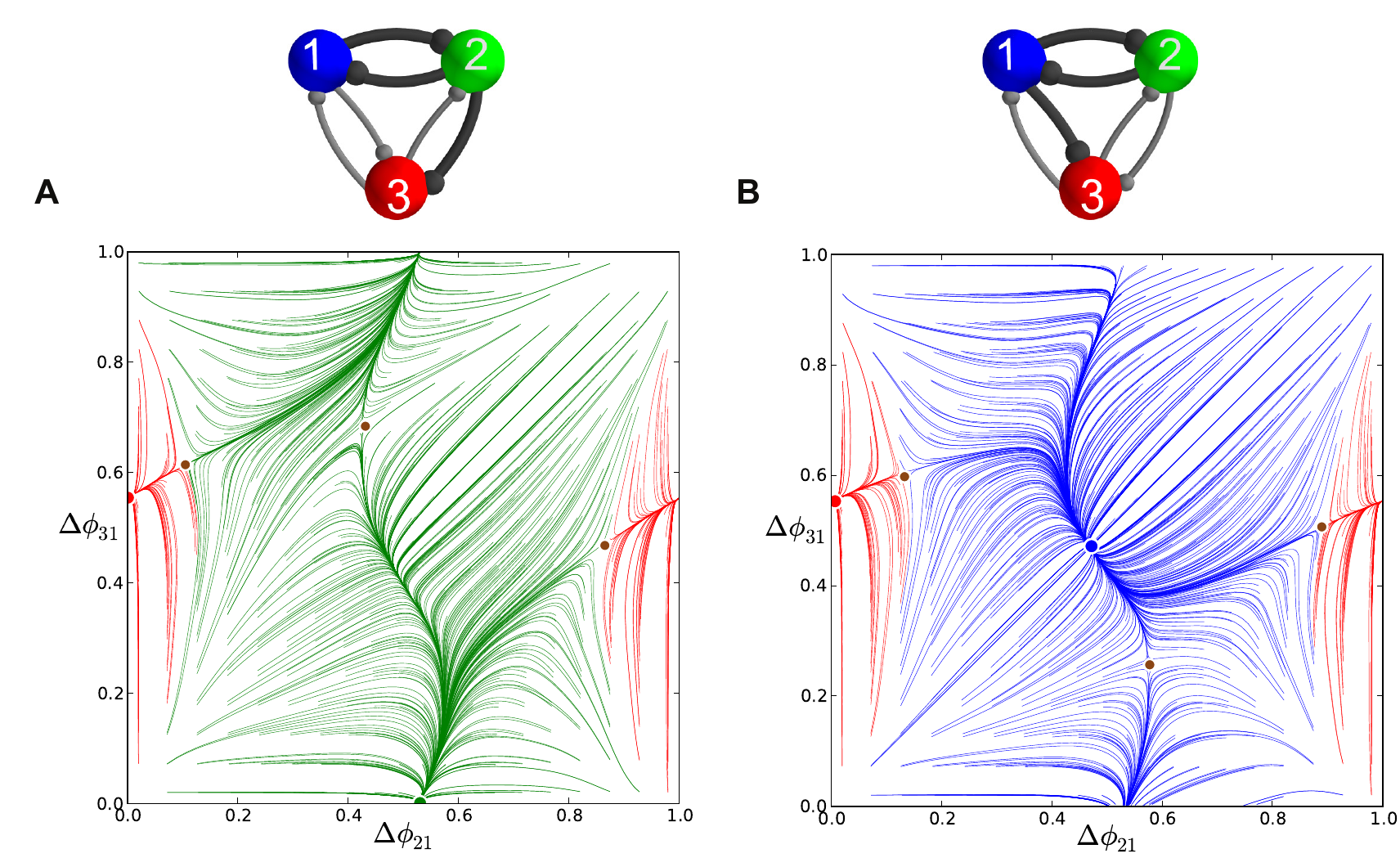}
\end{center}
\caption{ {\bf Motifs with the HCO asymmetrically inhibiting cell~3. }
(A) The phase lag map for the medium bursting motif at $\gsyn_{12}=\gsyn_{21}=\gsyn_{23} = 1.5
\gsyn$ generates two phase-locked bursting rhythms: a dominant $\TwoPerpOneThree$, due to a large attractor basin of
the green FP at $(\frac{1}{2},\,0)$, and $\ThreePerpOneTwo$ corresponding to the red attractor at $(0,\, \frac{1}{2})$ 
which has a smaller basin. (B) In the corresponding phase lag map at at $\gsyn_{12}=\gsyn_{21}=\gsyn_{13} = 1.5$, the stable FP at $(\frac{1}{2},\, \frac{1}{2})$ has a larger attractor basin  compared to that of the coexisting FP for cell~3 that leads the $\ThreePerpOneTwo$ rhythm.}\label{fig21}
\end{figure}

We now elucidate the issues involved in designing inhibitory motifs with predetermined bursting outcomes and how to control them. Let us revisit the  motif with a single HCO in Fig.~\ref{fig17}A. The map is depicted near the bifurcations that eliminate both blue and green FPs simultaneously as the coupling strength between cells~1 and 2 is increased. The corresponding saddle-node
bifurcations are each of co-dimension one, i.e. can be unfolded by a single parameter. This means that increasing either coupling parameter, $\gsyn_{12}$ or $\gsyn_{21}$,  makes the respective FP at $(\frac{1}{2},\,0)$ (green) or $\left (\frac{1}{2},\,\frac{1}{2} \right)$ (blue) disappear or re-emerge. This suggests alternative ways of perturbing the motif to get the desired outcome. For instance, in the motif with the HCO given by $\gsyn_{12}=\gsyn_{21} = 1.5\gsyn$, cell~2 can be made the strongest on the motif by increasing the outgoing inhibitory drive: $\gsyn_{23}=1.5\gsyn$.  The green FP at $\left (\frac{1}{2},\, 0 \right)$ in the corresponding map in Fig.~\ref{fig21}A has a largest attraction basin that guarantees the dominance of the  $\TwoPerpOneThree$-rhythm over the network. The map in Fig.~\ref{fig21}B has the basin of the blue FP at  $\left (\frac{1}{2},\, \frac{1}{2} \right)$ largest, after strengthening the coupling from cell~1 to 3 in the motif with two robust bursting outcomes: the $\OnePerpTwoThree$-rhythm  dominating over  the $\ThreePerpOneTwo$-rhythm corresponding to  the red FP at 
$\left (0,\, \frac{1}{2} \right)$ with a smaller basin formed by initial phases. 

%%%%% FIG 22 around here
%\vspace{0.2cm}{\it FIGURE 22 around here} \vspace{0.2cm}
\begin{figure}[h!]
\begin{center}
\includegraphics[width=0.55  \textwidth]{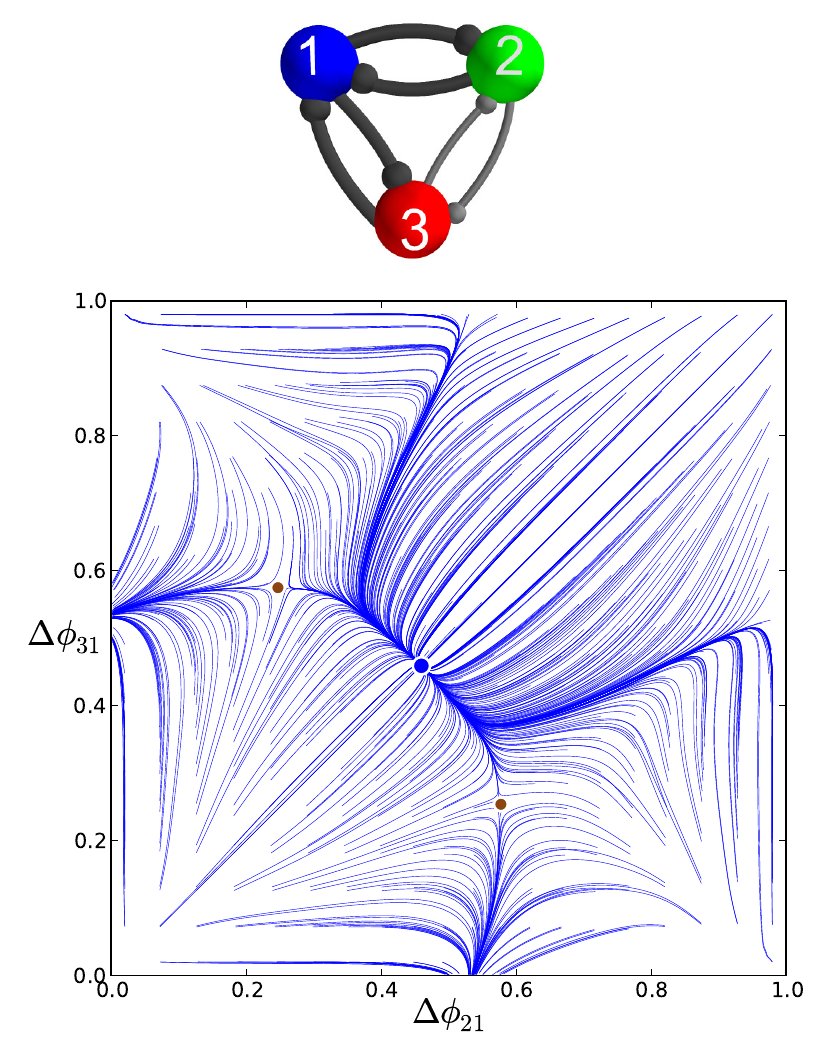}
\end{center}
\caption{ {\bf A motif with cell~1 leading in two half-center oscillators.}  The phase lag map at $\gsyn_{12}=\gsyn_{21} 1.5 \gsyn$ and $\gsyn_{13}=\gsyn_{31} = 1.5 \gsyn$ has a single phase-locked attractor -- the blue FP at $\left (\frac{1}{2},\, \frac{1}{2} \right)$  corresponding to the unique rhythm, $\OnePerpTwoThree$.}\label{fig22}
\end{figure}

The above configurations of the inhibitory motif are  bistable with two coexisting FPs: dominant blue (or green) with a large attractor basin and red with a smaller basin corresponding to the less robust $\ThreePerpOneTwo$ rhythm. To construct the monostable motif with the single rhythm, for example $\OnePerpTwoThree$,  cell~1 must be coupled \emph{reciprocally} stronger with cell~3 than cell 2. Such a motif has two HCOs that both contain cell~1 due to the strengthened pairs of synaptic connections: $\gsyn_{12}=\gsyn_{21}=1.5\gsyn$ and $\gsyn_{13}=\gsyn_{31}=1.5\gsyn$. The corresponding map for the phase lags is  shown in Fig.~\ref{fig22}.  The resulting map demonstrates that both the red and green FPs have been annihilated, as well as  the corresponding bursting rhythms. Note that the map  still has two saddle FPs in addition to the only attractor at $\left (\frac{1}{2},\,\frac{1}{2} \right )$. It is a feature of a map on a 2D torus that the number of FPs must be even, in general, for them to emerge and vanish through saddle-node bifurcations. Therefore, the map must possess another hyperbolic FP. This point resides near the origin where all three cells burst synchronously, which we consider next.

%%%%%%%%%%

\subsubsection*{Fine structure near the origin}

A common misconception concerning modeling studies of coupled cells is that fast, non-delayed inhibitory synapses always foster anti-phase dynamics over unstable in-phase bursting. While being true in general for simple relaxation oscillators, interactions of bursting cells can be incomparably more complex even in small networks including HCOs with fast inhibitory coupling \cite{prl08,pre2010}. It was shown in \cite{pre2012} that overlapped bursters can reciprocally synchronize each other in multiple, less robust, phase-locked states due to spike interactions. Furthermore, the number of such synchronous steady states is correlated with the number of spikes within the overlapped bursts. 

\begin{figure}[h!]
\begin{center}
\includegraphics[width=0.55  \textwidth]{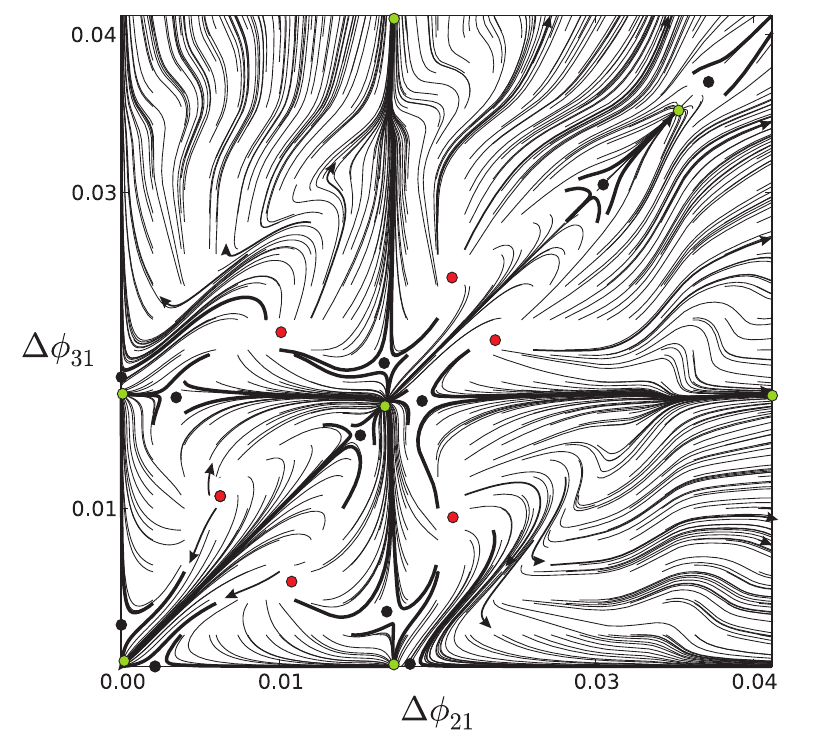}
\end{center}
\caption{ {\bf Fine dynamical structure near the origin $\Delta \phi_{21}=\Delta\phi_{31}=0$ of the phase lag map.}  Green, red, and black dots denote stable, repelling, and saddle FPs (resp.) in the vicinity of the origin, corresponding to all three cells almost synchronized in the homogenous medium-bursting motif. Globally, at a larger scale, the origin appears unstable.}\label{fig23}
\end{figure}

To explore the dual role of inhibition, we now explore nearly synchronous bursting in all three cells
of the homogeneous, medium DC motif. Because synchronous steady states are due to spike interactions, we restrict
the consideration to a relatively small positive vicinity of the synchronous state, $\Delta \phi_{21}=\Delta\phi_{31}=0$, in $\Pi$. A magnified portion of the map is shown in Fig.~\ref{fig23}, where green, red and black dots denote the locations of the stable, repelling and saddle (threshold) phase locked states (respectively) for the nearly synchronized bursting outcomes. The map reveals that several overlapping burst patterns can occur where either cell spikes slightly in advance or delayed compared to the reference cell.  Unstable FPs surround the outer part of this small region of the map make the origin repelling in the map on the global scale.

%%%%% FIG 23 around here
%\vspace{0.2cm}{\it FIGURE 23 around here} \vspace{0.2cm}

%%%%%%%%%%%%%%%%%%%%%%%%%%%%%%%%%%%%

\subsection*{Excitatory motifs}\label{excit}

In this section, we discuss a variation of a homogeneous 3-cell motif with short, 25\% DC at $\vks =-0.01895\mathrm{V}$, will all three excitatory synaptic connections.  The synaptic current is again given through the FTM paradigm: $I_{\rm syn}= \gsyn (E_{\rm syn} -V_{\mathrm{post}}) \Gamma (V_{\mathrm{pre}}-\Theta_{\rm syn})$. The synapses are made more excitatory by increasing the synaptic reversal potential, $E_{\rm syn}$,  from  $-0.0625\mathrm{V}$ (corresponding to the inhibitory case) to $0.0\mathrm{V}$. $E_{\rm syn} = 0$ guarantees that the voltages of all the cells remain  below the reversal potential, \emph{on average}, over the bursting period. In the excitatory motif, whenever the advanced cell initiatives a new bursting cycle, the synaptic current raises the voltages of post-synaptic cells, thus making it follow the pre-synaptic one,  at
the hyperpolarized knee-point on the quiescent manifold (Fig.~\ref{fig1}B).

%%%%% FIGURE 24 around here
%\vspace{0.2cm}{\it FIGURE 24 around here} \vspace{0.2cm}
\begin{figure}[h!]
\begin{center}
\includegraphics[width=0.55  \textwidth]{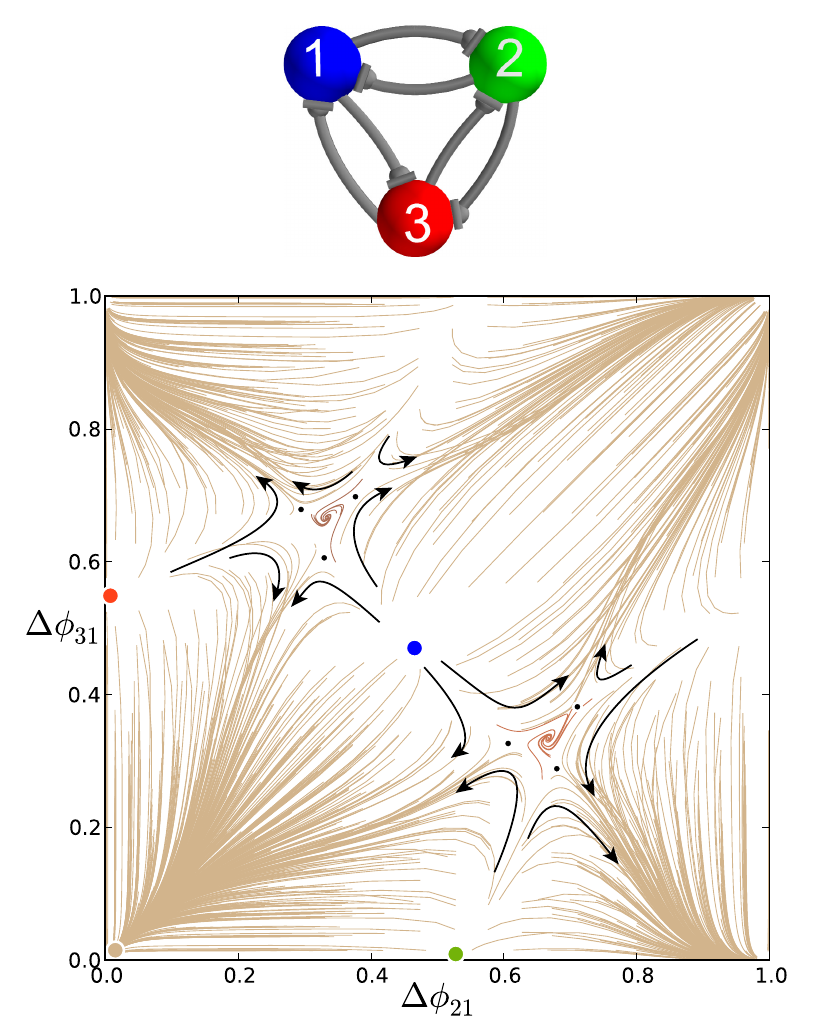}
\end{center}
\caption{ {\bf Phase lag map for the excitatory, weakly coupled, homogeneous motif with short duty cycle.}  With $\gsyn=5\times 10^{-4}$, this map has a dominant attractor at the origin that corresponds to synchronous bursting. Also depicted are three repelling FPs  (blue, red and green) at   $\left ( \Delta \phi_{21},  \Delta \phi_{31} \right) = \left (\frac{1}{2},\frac{1}{2} \right)$, $\left ( 0,\,\frac{1}{2}\right) $, and $\left (\frac{1}{2},\,0 \right)$, as well as stable FPs at  $ \left (\frac{2}{3},\frac{1}{3} \right)$ and $\left(\frac{1}{3},\frac{2}{3} \right)$ with small attractor basins, corresponding to  traveling waves, co-existing with the synchronous bursting.}\label{fig24}
\end{figure}

Figure~\ref{fig6} shows the phase lag map for the original inhibitory motif with three stable FPs  (shown in blue, red and green) at  $\left (\Delta \phi_{21},\, \Delta \phi_{31}\right ) = \left (  \frac{1}{2},\,\frac{1}{2} \right )$, $ \left (0,\,\frac{1}{2} \right)$, $\left ( \frac{1}{2},\,0 \right )$ and two unstable FPs (dark dots) at $\left(\frac{2}{3},\frac{1}{3} \right )$ and $\left( \frac{1}{3},\frac{2}{3} \right)$. The attractor basins of three stable FPs are separated by the separatrices of six saddle FPs (smaller dots). A small area around the origin is globally repelling. This motif can stably produce three coexisting patterns in which either cell bursts in anti-phase with the two remaining in-phase.    

It is often presumed in neuroscience that excitation acts symmetrically opposite to inhibition in most cases, i.e. wherever inhibition tends to break synchrony, excitation fosters it. Figure~\ref{fig24} supports this assertion for this particular kind of network and coupling. It depicts the map corresponding to the homogeneous 3-cell motif with reciprocally excitatory connections for same short, 25\% DC.

%%%%% FIGURE 25 around here
%\vspace{0.2cm}{\it FIGURE 25 around here} \vspace{0.2cm}
\begin{figure}[h!]
\begin{center}
\includegraphics[width=0.99  \textwidth]{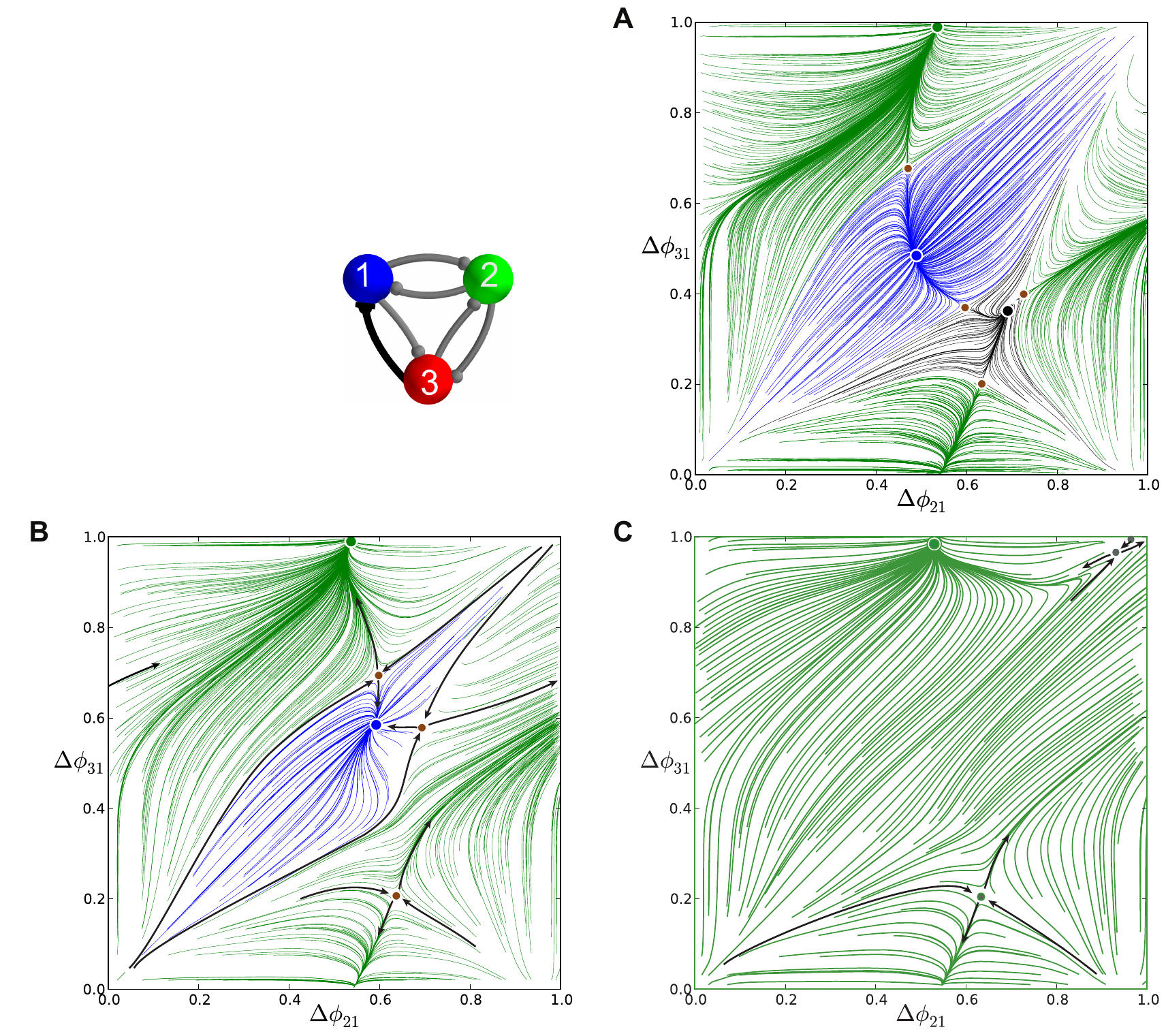}
\end{center}
\caption{ {\bf Phase lag maps for the mixed homogeneous motif with medium duty cycle as reversal potential varies.} With $\gsyn=0.0005$, maps are depicted for $E^{31}_{\rm syn}=-0.050$ in (A), $-0.030$ in (B) and $-0.020\mathrm{V}$ in (C). (A) Increasing  $E_{\rm syn}^{31}$ causes two saddle-node bifurcations: one breaks the clockwise rotational symmetry and annihilates the corresponding FP at $\left (\frac{1}{3},\frac{2}{3} \right )$, while the other annihilates the stable red point at  $(0,\,\frac{1}{2})$. (B) This widens the basins of the blue and green stable FPs at $ \left  (\frac{1}{2},\frac{1}{2} \right) $ and $ \left (\frac{1}{2},\,0 \right )$. Further raising $E_{\rm syn}^{31}$ eliminates the attractor basin for the black FP at  
$ (\frac{2}{3},\frac{1}{3})$ in (B), and finally the blue FP along with the $\OnePerpTwoThree$- rhythm in (C). Black-labeled trajectories indicate the direction field on the torus and the separatrices of saddles.}\label{fig25}
\end{figure}

Compared to the map for the inhibitory motif, the map for the homogeneously excitatory motif is the  inverse:  
$ \Pi^{-1}:~\left (\Delta \phi_{21}^{(n+1)},\, \Delta \phi_{31}^{(n+1)} \right)  \to \left ( \Delta \phi_{21}^{(n)},\, \Delta \phi_{31}^{(n)} \right )$; here the inverse is the forward map in discrete backward time. As such, the FPs at $ \left (\frac{1}{3},\frac{2}{3} \right ) $ and $\left (\frac{2}{3},\frac{1}{3} \right )$, which used to be repelling in the inhibitory case, become attracting but with  smaller basins. This means that the motif can generate traveling waves, albeit with low probability. Meanwhile, the FPs colored blue, green and red, are now repellers, and hence none of the pacemaker rhythms can occur. Reversing the stability does not change the topological type of the six saddles, but their stable and unstable  separatrices are reversed. The dominant attractor of the map is now located at the origin, to which nearly all transient trajectories converge. This implies that the reciprocally excitatory motif, whether homogeneous or heterogeneous, will exhibit stable synchronous bursting with all three cells oscillating in-phase.    
  
%%%%% FIGURE 26 around here
%\vspace{0.2cm}{\it FIGURE 26 around here} \vspace{0.2cm}

\begin{figure}[h!]
\begin{center}
\includegraphics[width=0.99  \textwidth]{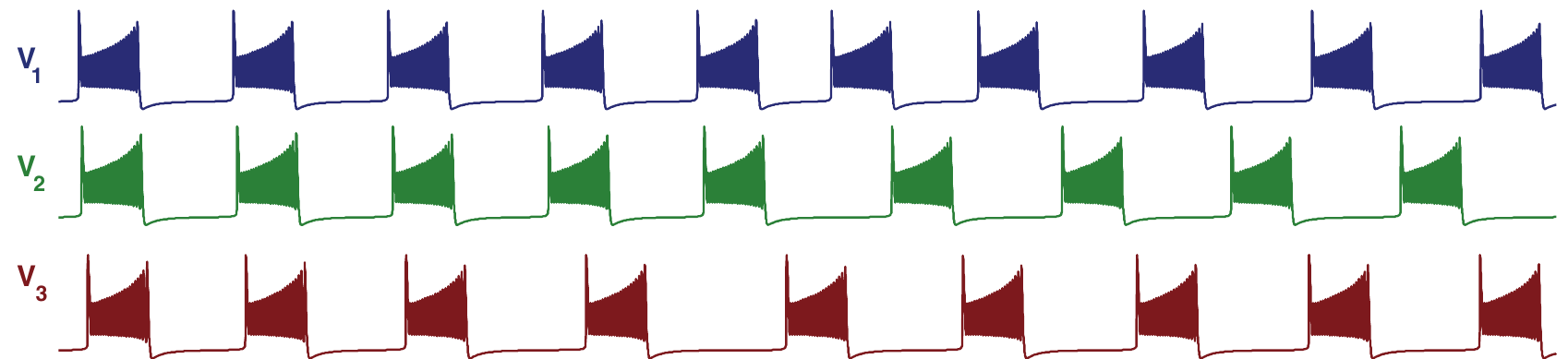}
\end{center}
\caption{ {\bf Transient voltage traces converging to the $\TwoPerpOneThree$-rhythm 
generated by a mixed motif.} Here, $E^{31}_{\rm syn}=-0.020\mathrm{V}$, which corresponds to the phase lag map having a single attractor at the green FP in Fig.~\ref{fig24}D; here $\gsyn=0.005$ was used  for the sake of illustration.}\label{fig26}
\end{figure}
%%%%%%%%%%%%%%%%%%%%%%%%%%%%%%%%%%%%%%%%%%%%

\subsection*{Mixed motifs}\label{mixed}

Here we discuss two intermediate configurations of mixed motifs having both inhibitory and excitatory connections. First, we consider the motif with a single excitatory connection from cell~3 to~1. Its coupling strength is regulated by the level of the synaptic reversal potential, $E_{\rm syn}^{31}$. Figure~\ref{fig25} 
depicts three phase lag maps for the motif with  $E_{\rm syn}$ being increased from $-0.050$, $-0.030$ through $0.0\mathrm{V}$.

%%%%% FIGURE 27 about here

%\vspace{0.2cm}{\it FIGURE 27 about here}
\begin{figure}[h!]
\begin{center}
\includegraphics[width=0.99  \textwidth]{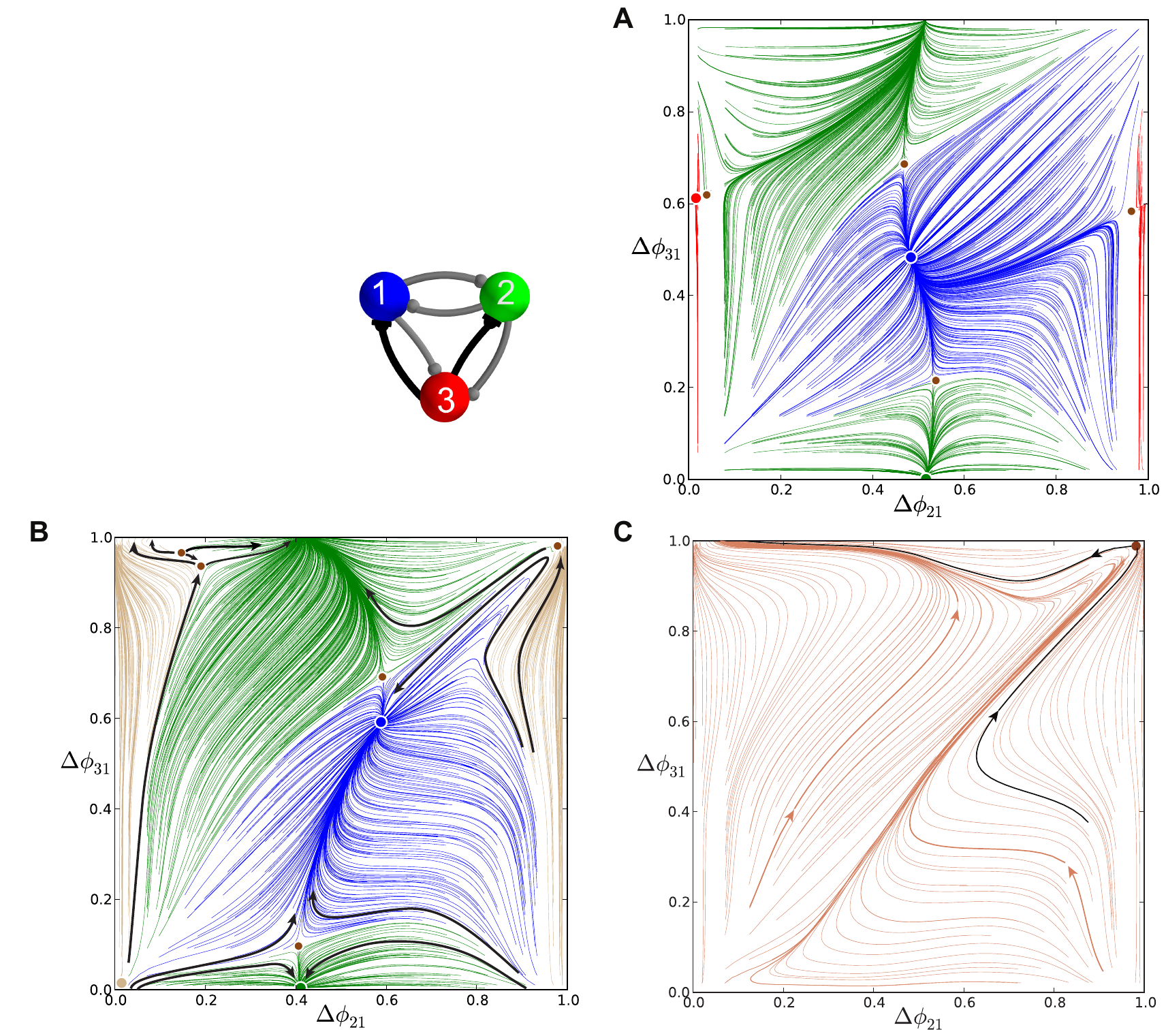}
\end{center}
\caption{ {\bf Phase lag maps for the mixed motifs as an excitatory reversal potential varies.} Here, we choose values $-0.050$, $-0.030$ and $-0.010\mathrm{V}$, for the reversal potential, $E_{\rm syn}$ in the two excitatory connections originating from cell~3.  (A) Increasing $E_{\rm syn}$ causes two saddle-node bifurcations, and breaks the rotational symmetries and hence annihilates the FPs at $ \left (\frac{2}{3},\frac{1}{3} \right )$ and $\left ( \frac{1}{3},\frac{2}{3} \right )$.  This widens the basins of the blue and green stable FPs at $ \left (\frac{1}{2},\frac{1}{2} \right )$ and  $\left (\frac{1}{2},\, 0 \right )$, and shrinks that of the red stable FP at $ \left (0,\,\frac{1}{2} \right )$. (B) Making two connections more excitatory produces a closed heteroclinic connection between the remaining FPs, which becomes  a stable invariant circle wrapped around the torus (inset~C). Black-labeled trajectories
indicate the direction field on the torus and the separatrices of saddles.}\label{fig27}
\end{figure}

Initially, an increase in $E_{\rm syn}^{31}$ gives rise to two saddle-node bifurcations in the motif (Fig.~\ref{fig25}A): the first one breaks the clockwise rotational symmetry and hence annihilates the stable FP at $ \left (\frac{1}{3},\frac{2}{3}\right )$. The second bifurcation annihilates the stable red point at  $\left  (0,\,\frac{1}{2} \right)$, because cell~3,  inhibiting~2 and exciting~1, cannot hold both of them at the hyperpolarized quiescent state
to generate the $\ThreePerpOneTwo$-rhythm as it promotes burst initiation in cell 1 following those in cell 3. On the contrary, excitation applied to cell~1 forces it to follow cell~3   after a short delay in the burst initiation. As the result, the disappearance of the $\ThreePerpOneTwo$-rhythm promotes the $\TwoPerpOneThree$-rhythm and an increase of the attractor basin of the green FP.  

Initial elevations of the level of $E_{\rm syn}^{31}$ keep the other three FPs intact, while widening the basins of the blue and green stable FPs at $ (\frac{1}{2},\frac{1}{2})$ and $(\frac{1}{2},\,0)$. Further increasing $E_{\rm syn}^{31}$  increases the duty cycle of the blue cell by extending its active bursting phase. Consequently, the counter-clockwise ring no longer contains identical cells that could orchestrate the $\OneThreeTwo$ pattern. This patten  is eliminated with the disappearance of the corresponding FP at  $(\frac{2}{3},\frac{1}{3})$ through a merger with a saddle. The map now has two persistent attractors, blue and green, as shown in Fig.~\ref{fig25}B. With $E_{\rm syn}^{31}$ increased still further, the blue cell~1 receives strongly unbalanced input: larger excitation influx from the postsynaptic cell~3 and an inhibitory drive from cell~2, acting oppositely. This unbalanced input increases the active phase of bursting of cell 1 and hence its duty cycle and period, and hence breaks cell~1's ability to robustly maintain the $\OnePerpTwoThree$-rhythm by evenly inhibiting the pots-synaptic cells 1 and 2 of the same period. In $\Pi$, this results in the shrinking of the attractor basin of the blue FP, whereas the basin of the dominating green FP widens. By setting  $E_{\rm syn}^{31}=0.0\mathrm{V}$, the resulting strong imbalance between excitation and inhibition onto cell~1 makes the $\OnePerpTwoThree$-rhythm impossible to occur in the network and the corresponding FP at $ \left (\frac{1}{2},\frac{1}{2} \right )$ disappears in the map. After this last saddle-node bifurcation, the map has a unique attractor in the green FP. Eventually, regardless of initial phases, the synergetic interaction of inhibitory and excitatory cells in this mixed motif will give rise to the $\TwoPerpOneThree$-rhythm led by cell~2 (Fig.~\ref{fig26}).

Finally, we consider the mixed motif with two excitatory connections originating from cell~3. Figure~\ref{fig27} depicts the transformations of the $\Pi$ as the reversal potentials, $E^{31}_{\rm syn}$ and $E^{32}_{\rm syn}$, are increased simultaneously from $-0.050$, $-0.030$ through to $-0.020\mathrm{V}$. The increase makes postsynaptic cells~1 and 2
more excited compared to cell~3, which consequently receives a longer duration of inhibition.

As in the previous case, increasing  $E^{31}_{\rm syn}=E^{32}_{\rm syn}=-0.050\mathrm{V}$ breaks both rotational symmetries, which is accompanied by the disappearance of the corresponding, counter-clockwise  and  clockwise, FPs. The increased excitability of cells~1 and 2 initiates the active bursting states of those cells soon after that of cell~3 in each cycle.  Thus, the  $\ThreePerpOneTwo$-rhythm led by cell~3 is less likely to occur compared to the $\OnePerpTwoThree$ and $\TwoPerpOneThree$- rhythms. The corresponding (red) FP at $\left (0,\,\frac{1}{2} \right )$ loses the attractor basin in the map and then disappears, following the FPs for the traveling waves, at $-0.030\mathrm{V}$, see Fig.~\ref{fig27}B. After that, the blue and green stable points have equal attractor basins corresponding to equal chances of the emergence of the phase-locked rhythms  $\OnePerpTwoThree$  and $\TwoPerpOneThree$.  Examination of the map suggests that besides these phase-locked rhythms, the motif can generate long transients with episodes that resemble those of the $\ThreePerpOneTwo$-rhythm  transitioning back and forth with in-phase bursting. Such transients are due to regions of the map that are constrained by the incoming separatrices of the remaining saddles, which are forced to curve in complex ways to embed onto the torus with two attractors and an unstable origin. Note that the origin may not longer be a repeller as a whole but has a saddle structure, because it is no longer associated with synchronous bursting.

Setting the excitatory reversal potential to zero annihilates the two remaining phase-locked states (blue and green FPs) in two simultaneous bifurcations. As with the case of the inhibitory motif with the HCO (compare with Fig.~\ref{fig18}), these global bifurcations underlie the formation of a closed heteroclinic connection between the saddle-node points at the critical moment. These connections transform into an invariant circle that wraps around the torus. Having settled onto the invariant curve that zigzags over the torus,  the network will generate long recurrent patterns consisting of three transient episodes: namely, in-phase bursting that transitions to the $\OnePerpTwoThree$-rhythm, which transitions back to in-phase bursting, then transitioning to the $\TwoPerpOneThree$-rhythm, which then returns to the in-phase bursting, and so forth.

At higher values of reversal potentials, excitation overpowers inhibition and this mixed motif fully becomes the excitatory motif with the single, all-synchronous, bursting rhythm forced by the driving cell~3. In the corresponding return map, this rhythm occurs after the  invariant curve terminates at a homoclinic connection to a saddle-node near the origin so that the origin becomes a global attractor again.

\subsection*{Gap junction in an inhibitory motif}

\begin{figure}[h!]
\begin{center}
\includegraphics[width=0.99  \textwidth]{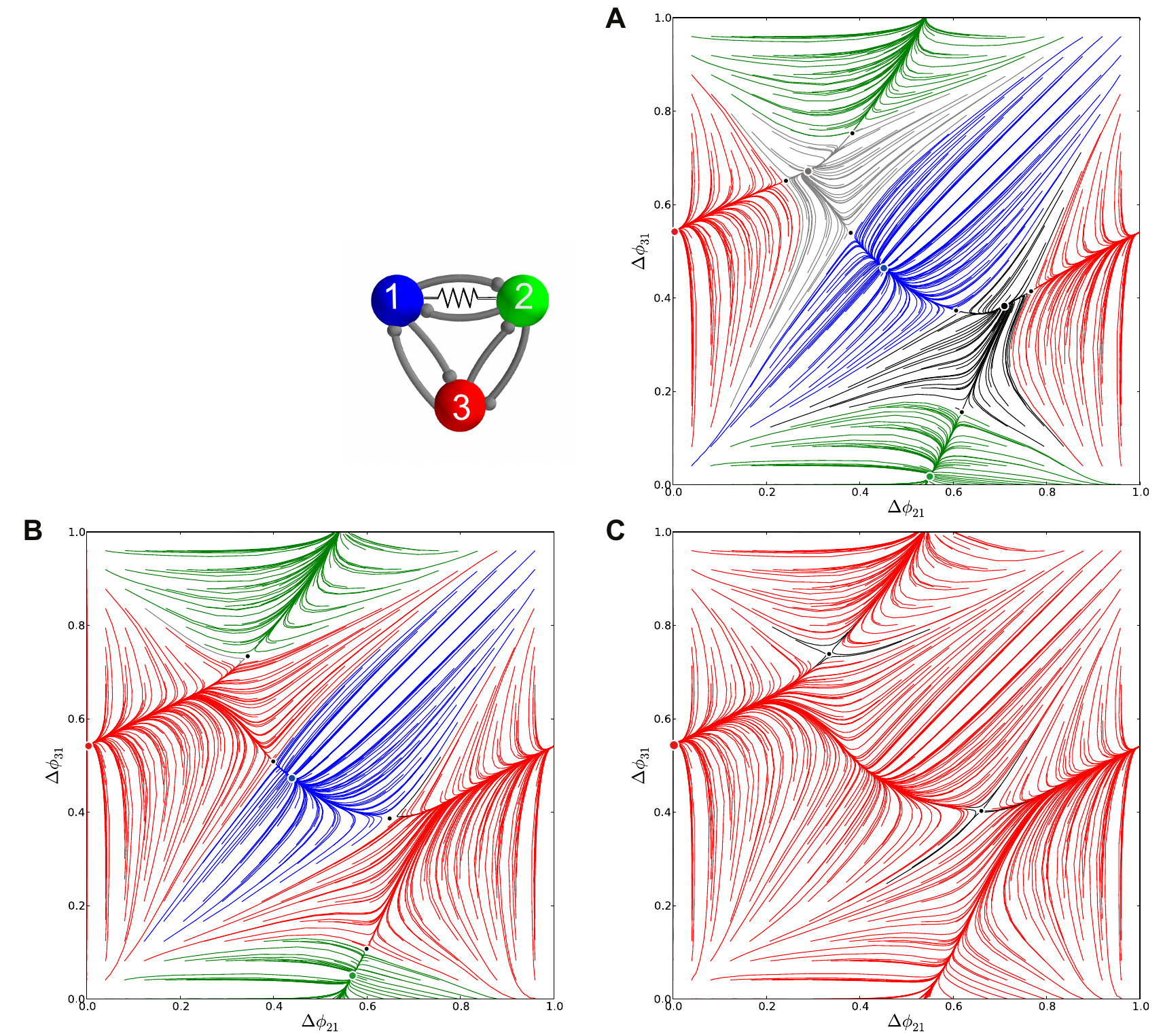}
\end{center}
\caption{ {\bf Transformation stages of the phase lag map for the inhibitory motif with a single gap junction between cells 1 and 2.} Increasing the electrical coupling strength from $10^{-4}$ through $3\times 10^{-4}$ transforms the multistable motif into a dedicated one by eliminating first the FPs corresponding to the traveling waves, and next the green and blue FPs at the same time as the gap junction is bi-directional. Eventually, the red FP $\left (0,\,\frac{1}{2} \right )$, corresponding to the single $\ThreePerpOneTwo$-rhythm led by cell 3 in the motif with the gap junction uniting bursting cells 1 and 2, becomes the global attractor of the map.} \label{fig28} 
\end{figure}

Finally, let us consider the role of a single electrical synapse though a gap junction between cells 1 and 2 in the inhibitory motif (Fig.~\ref{fig1}A). The difference in the membrane potentials gives rise to a directional ohmic current described by $I_{\rm el}=g_{\rm el}(V_2 -V_1)$ in the model~(\ref{eq1}). Figure~\ref{fig28} depicts the stages of transformation of the corresponding maps as $g_{\rm el}$ is increased from $10^{-4}$ through $3\times 10^{-4}$. 

The electrical coupling breaks down the rotational symmetry that causes the disappearance of the corresponding FPs at $\left (\frac{1}{3},\,\frac{2}{3} \right )$  and  $\left (\frac{2}{3},\,\frac{1}{3} \right )$. The disappearance of both FPs widens the attraction basin of the red FP at $\left (0,\,\frac{1}{2} \right )$ compared to the individual basins of the blue and green FPs. The bidirectional electrical coupling tends to equilibriate the membrane potentials of the connected cells, so that cell 1 and 2 are brought closer together to burst in synchrony, rather than in alternation. At intermediate values of $g_{\rm el}$, inhibitory coupling can still withstand the tendency to synchronize cells 1 and 2, while the red basin widens further due to shrinking basins of the blue and green FP. At larger values of  $g_{\rm el}$, synchrony of bursting cells 1 and 2 overpowers their reciprocal inhibition, and the motif generates only the $\ThreePerpOneTwo$-rhythm  regardless of the choice of initial phases. Thus, we find that motifs with strong electrical synapses describe  a dedicated rather than a multifunctional CPG.    

%%%% FIGURE 28 about here

%\vspace{0.2cm}{\it FIGURE 28 about here}

%%%%%%%%%%%%%%%%%%%%%%%%%%%%%%%%%%%%%%%%

\section*{Discussion}

Our new computational technique reduces the dynamics of a 9-dimensional network motif of three cells to the analysis of the equationless 2D maps for the phase lags between the bursting cells. With this technique, we demonstrated that a reciprocally inhibitory network can be multistable, i.e. can generate several
distinct polyrhythmic bursting patterns. We studied both homogeneous and non-homogeneous coupling scenarios as well as mixtures of inhibition, excitation and electrical coupling. We showed that the observable rhythms of the 3-cell motif are determined not only by symmetry considerations but also by the duty cycle, which serves as an order parameter for bursting networks. The knowledge of the existence, stability and possible bifurcations of polyrhythms in this 9D motif composed of the interneuron models
is vital for derivations of reduced, phenomenologically accurate phase models for non-homogeneous biological CPGs with inhibitory synapses.

The idea underlying our computational tool is inspired by features of a ``wet lab'' experimental setup. As such, it requires only the voltage recording from the model cells and does not explicitly rely on the gating variables. We intentionally choose the phases based on voltage as often this the only experimentally measurable variable. Moreover, as with real experiments, we can control the initial phase distribution by releasing the interneurons from inhibition at specific times relative to the reference neuron.

%%% SUMMARY OF RESULTS 
\subsubsection*{Summary and interpretation of main results}

In the Homogeneous Inhibitory Motifs section, we showed that a weakly coupled, homogeneous motif comprised of three bursting interneurons with fast reciprocally inhibitory synapses can produce a variable number of polyrhythms, depending on the duty cycle of the individual components. The phase lag maps are \emph{de facto} proof of the robust occurrence of the corresponding rhythmic outcomes generated by such a motif. While the occurrence  of some rhythms, such as $\OneTwoThree$ and $\OneThreeTwo$, in a 3-cell motif can hypothetically be deduced using symmetry arguments; the existence and robustness of the rhythms can be only verified by accurate computations of the corresponding return maps. Moreover, the observability of these rhythms in even the homogeneous motifs, and  the stability of the FPs, are both closely linked to the temporal properties of the bursts.

Recall that the inhibitory current shifts the post-synaptic cell closer to the boundary or can even move it over the boundary into silence while the pre-synaptic cell remains actively spiking (Fig.~\ref{fig3}). In terms of dynamical systems theory, this means that the perturbed driven system closes the gap between the hyperpolarized fold
and the slow nullcline $m^\prime_{\rm K2}=0$, eventually causing the emergence of a stable equilibrium state on
the quiescent branch. As such, the homogeneous network produces three pacemaker rhythms, $\ThreePerpOneTwo$, $\TwoPerpOneThree$, and $\OnePerpTwoThree$ -- the only rhythms available in the short motif. These strongly synchronized activities imply fast convergence to the phase locked states because of the emergent
equilibrium state near the closed gap: compare the time spans in Figs.~\ref{fig6}A and \ref{fig9}A.

The gap never gets closed by weak inhibition in the long bursting motif as the individual cells have
initially remained far enough from the boundary separating the bursting and hyperpolarized quiescence. As the result, this motif can only effectively produce two possible bursting outcomes: the clockwise 
$\OneTwoThree$  and counter-clockwise $\OneThreeTwo$ traveling waves. This bistability results from a weaker form of synchronization, which is confirmed by the rate of convergence to the FPs. This will not be the case when the inhibition becomes stronger due to increasing the nominal coupling strength, $g_\mathrm{syn}$.

The intermediate case of the medium motif is far from the above extremes and benefits from a natural optimization between the coupling strength, initial phase distributions, and the spatio-temporal
characteristics of unperturbed and perturbed bursting. One such characteristic is the slow passage through the ghost of the stable equilibrium state that guarantees the robust synchrony in the short bursting motif. As the result, the motif can produce five stable bursting rhythms:
the anti-phase $\ThreePerpOneTwo$, $\TwoPerpOneThree$, and $\OnePerpTwoThree$; and the clockwise $\OneTwoThree$  and counter-clockwise  $\OneThreeTwo$ traveling waves.

In the Asymmetric Inhibitory Motifs section, we described a number of generic bifurcations of the five original five polyrhythms that can occur
in the homogeneous, reciprocally inhibitory motif with the medium duty cycle.  We revealed the basic principles of transformations of such a multi-functional network into ones with fewer rhythms or even with a single pattern bursting pattern.

%%%%%%%% FIGURE 29 ABOUT HERE
%\vspace{0.2cm}{\it FIGURE 29 ABOUT HERE}
\begin{figure}[h!]
\begin{center}
\includegraphics[width=0.99  \textwidth]{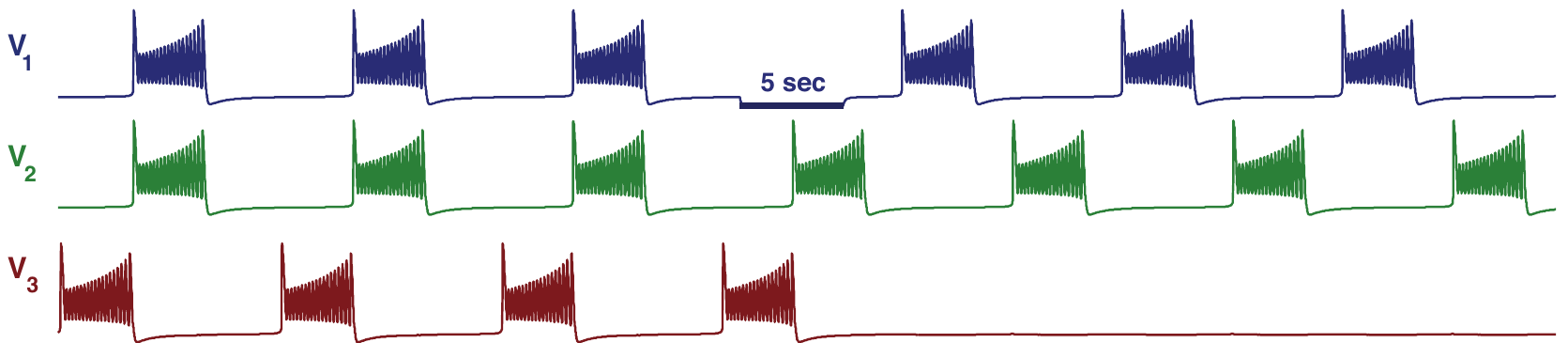}
\end{center}
\caption{ {\bf \emph{Sudden death} of bursting in cell~3 after application of an inhibitory stimulus to cell~1.} An inhibitory stimulus causes the switch from the $\ThreePerpOneTwo$ rhythm, led by cell~3, to a pattern where it is forced to become hyperpolarized quiescent. This state is due to cell~3 receiving continuous inhibition by the half-center oscillator formed by cells 1 and 2 in an  asymmetric motif at $\gsyn_{13}=\gsyn_{23}=6 \gsyn$ and $\gsyn_{31}=\gsyn_{32}=0.4\gsyn$.}\label{fig29}
\end{figure}

The rhythmic outcomes of the CPG do not always have to involve phase locking, as there can be a stable pattern of \emph{phase slipping} bursts that have time varying phase lags. 

While each rhythm remains robust with respect to variations of the coupling connections,
one can still make the network switch between them by applying an external pulse to the targeted cell that, upon release, appropriately changes the relative phases of the cells, as demonstrated in Fig.~\ref{fig10} for the medium bursting motif. In terms of the Poincar\'e return maps for the phase lag, switching between rhythms is interpreted as switching between the corresponding attractor basins of FPs or invariant circles. This causes the state of the network to ``jump'' over the separating threshold defined by a saddle (more precisely, over the incoming separatrices of the saddle). We stress, however, that the choice of timing in the suppression of a targeted cell to effectively switch between these polyrhythms is not intuitive and requires a detailed understanding of the underlying dynamics.

Note that, although there are alternative ways of creating the 3-cell reciprocally inhibitory motif with predetermined outcomes,
the fundamental principles are universal: the pre-synaptic cell that produces stronger inhibition gains the larger attractor basin and therefore the corresponding rhythm led by this cell becomes predominant. In particular, a sufficient increase (or decrease) of the coupling strength of a single connection can break the intrinsic rotational symmetry of the motif to remove the possibility of observing traveling wave patterns.

Of special interest are the motifs with phase-slipping patterns that have no dominating phase-locked states.
Such a pattern results from the synergetic interactions between all contributing cells, and is comprised of four transient episodes, but primarily marked by a continuous transition between two primary sub-rhythms: $\TwoPerpOneThree$ and $\OnePerpTwoThree$. Both competing sub-rhythms are equally possible, and none can prevail over the other without the weaker inhibiting cell~3 whose reciprocal connections change the existing balance by shifting the phase lags during all four episodes.

In all cases we have considered, inhibition was chosen weak enough to guarantee that  the post-synaptic cell remains in a bursting state even when its duty cycle decreases to  $20\%$  near the boundary between bursting and silence. This ensures that the phases of the cells, as well as the phase lags,  are well defined for the return maps.  However, this assumption may fail, for example when phase of a post-synaptic cell is not defined because the incoming synaptic inhibition makes it quiescent (Fig.~\ref{fig29}) \cite{Shilnikov2008b}. This leads to a phenomenon called \emph{sudden death} of bursting that occurs when a rhythmic leader cell (red cell~3 in this figure) becomes suppressed by the other two cells that form an anti-phase HCO, which alternately inhibit the post-synaptic cell. In our example, the outgoing inhibition from cell~3 is several times  weaker  than the inhibition that it receives from the HCO formed by cells~1 and 2: $\gsyn_{31}=\gsyn_{32}=0.4\gsyn$ and $\gsyn_{13}=\gsyn_{23}=6 \gsyn$ (this ratio does not always have to be as large when $\gsyn$ is increased).

It should be stressed that the sudden death of bursting co-exists with other bursting patterns in the motif at the same coupling parameters.    As such,  this example bears a close qualitative  resemblance to the experimental recordings from identified interneurons comprising the leech heart CPG in which the so-called switch interneuron alternately leads synchronous patterns and then becomes inactive during peristaltic patterns  \cite{1018167,11877527, 11877529}. Analogous reversions of direction in the blood circulation,  peristaltic and synchronous, are observed in the leech heartbeat CPG and its models  \cite{22329853,Calabrese01122011}. The contrast between the patterns is that switching appears to be autonomously periodic, i.e. without external stimuli, in a way similar to the phase-slipping pattern presented in Fig.~\ref{fig19}. 

Examination of the complex fine structure of the map near the origin reconfirms that fast, non-delayed inhibition can have stabilizing effects leading to the onset of several nearly synchronous bursting patterns with several overlapping spikes \cite{pre2012}. Such bursting patterns are less robust compared to those corresponding to FPs with large attractor basins in the phase lag maps.     

We showed in the Excitatory Motifs section that raising the synaptic reversal potential is equivalent to reversing the time in the inhibitory motif model. In the maps, this action makes attractors into repellers, while saddles remain saddles but have their incoming and outgoing directions reversed. To fully illustrate this phenomenon, we used the symmetric motif with the short duty cycle, in which the FPs for clock and counter-clockwise traveling waves are unstable. In the symmetric counterpart with excitatory synapses, these FPs become the attractors along with the dominant fixed  point at the origin corresponding to synchronous bursting. 

In the Mixed Motifs section we analyzed the transformation of the maps corresponding to the motifs with mixed, inhibitory and excitatory synapses. We showed step-by-step how the the multi-functional motif becomes a dedicated motif. The final example was the mixed motif with two excitatory connections. Unlike the former case, such a monostable motif possesses a bursting rhythm with time varying phase lags, which corresponds to a stable invariant curve wrapping around the 2D torus.      

%%%%%% END OF LONG RESULTS SUMMARY

\subsubsection*{Conclusions and future work}

We emphasize that a highly detailed examination of the occurrence and robustness of bursting patterns in the 3-cell motifs would be impossible  without the reduction of the complex 9D network model with six algebraic equations for chemical synapses to the 2D maps and the numerical bifurcation analysis of FPs, invariant circles, homoclinic structures in it. Recall that the dimension of the map is determined by the number of the nodes in the network, not the number of differential equations per a synapse and per a neuron, which can be much greater. With the aid of our computational tools we were able to identify even some exotic bifurcations from the dynamical system theory like the Cherry flow (Fig.~\ref{fig27}D) on the torus \cite{books} occurring in this neural network. High accuracy of numerical simulations is required in our analysis. This involves at least a $40\times 40$ mesh of initial conditions run for 100 bursting cycles to generate a single map and identify its structure. This takes at least two hours on a multi-core CPU workstation, but future work will take advantage of parallel computing architectures.

\begin{figure}[h!]
\begin{center}
\includegraphics[width=0.55  \textwidth]{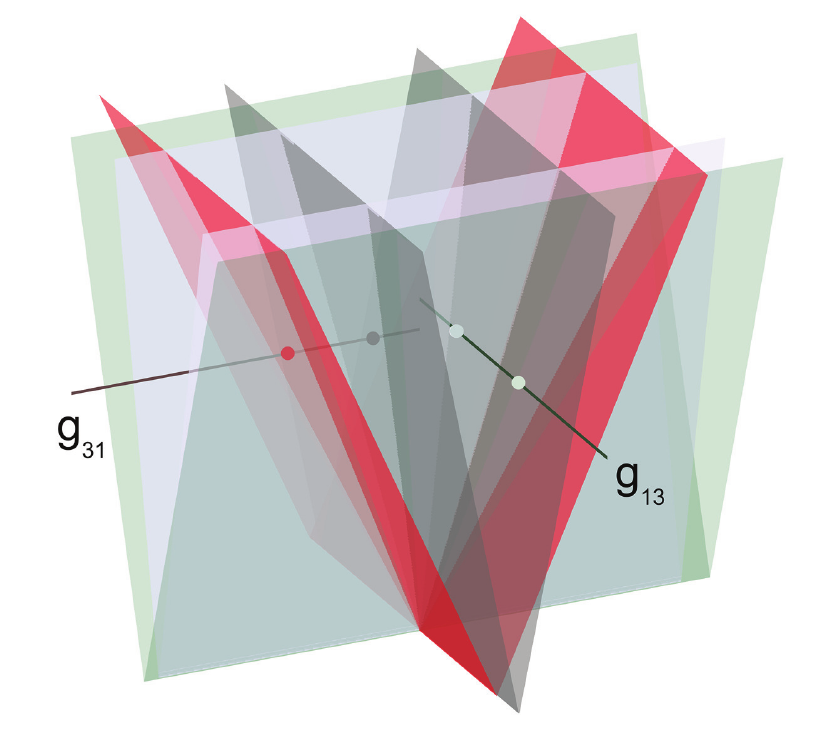}
\end{center}
\caption{{\bf Sketch of nested synchronization zones in the parameter space of the network.} The boundaries of the zones are crossed when a single bifurcation parameter, here the coupling strength, $\gsyn_{ij}$, is varied.  Such borders, on which saddle-node bifurcations occur, bound the existence regions of stable FPs corresponding to various phase-locked bursting  patterns of the CPG. }\label{fig30}
\end{figure}

A stable FP of the map corresponding to a robust bursting pattern with specific phase lags is also structurally stable, i.e. persists under particular variations of coupling parameters. So by varying the given parameter(s) we can evaluate the boundaries and region of its existence and. A boundary of the region corresponds to a bifurcation, which cab be either a saddle-node, through which the FP her vanishes, or only looses the stability as in the Andronov-Hopf case. In terms of the classical synchronization theory \cite{shilnikov2004some}, the FP is mathematical representation of a phase-locked state, in the given context, it is  a bursting pattern with some constant phase lags between the cells of the network. Its existence/stability region is called a synchronization zone, as the bursting cells become synchronized in a certain order -- pattern. 
 Using the qualitative tools of the bifurcation theory, one can quantitatively estimate the size of synchronization zones and determine their boundaries in the parameter space of the dynamical system representing the 3-cell CPG network under consideration. Figure~\ref{fig30} sketches how such zones can be embedded in such a parameter space.  Some zones are nested within each other as changing of a single parameter of the network can consequently cause a cascade of saddle-node bifurcations in the map, for instance, such as ones shown in Figs.~\ref{fig18} and \ref{fig19}. Given the number of the reciprocal synapses in the motif, the parameter space is at least six-dimensional, which presents quite a challenge for detailed bifurcation examinations. We have demonstrated though that several  inhibitory configurations of the 3-cell motif generate phase lag map of the {\em qualitatively} same structure, 
see exemplary Fig.~\ref{fig18}.  This observation provides underlying foundations for highly effective reduction tools for studies of multi-component neural networks. It implies that variations of different coupling parameters make the network undergo same bifurcations while crossing transversally the corresponding  bifurcation boundaries in the parameter space.                 
This {\em de-facto} proves again that without the phase lag maps, it would be impossible to claim and understand why  two distinct network configurations produce same rhythmic outcomes.

%%%%%% FIGURE 30 ABOUT HERE

%\vspace{0.2cm}{\it FIGURE 30 ABOUT HERE}
%\vspace{0.2cm}

In general, our insights allow us to predict both quantitative and qualitative transformations of the observed patterns as the network configurations are altered or the network states are perturbed dynamically \cite{burylko2008}. The nature of these transformations provides a set of novel hypotheses for biophysical mechanisms about the control and modulation of rhythmic activity. A powerful aspect to our analytical technique is that it does not require knowledge of the equations that model the system (even though we happened to have them here). The computational tools help us explain the fundamental dynamical mechanisms underlying the rhythmogenesis in plausible models of CPG networks derived from neurophysiological experiments \cite{chaos_melibe}.  Thus, we believe that have developed a universal approach to studying both detailed and phenomenological models that is also applicable to a variety of rhythmic biological phenomena beyond motor control.

%%%%%%%%%%%%%%%%%%%%%%%%%%%%%%%%%%%%%%%%%%%%%%%%%%%%%%

\section*{Models and Numerical Methods}

We study CPG network motifs comprised of three cells coupled reciprocally by non-delayed, fast chemical synapses and with weak coupling strengths.  The time evolution of the membrane potential, $V$, of each neuron is modeled using the framework of the Hodgkin-Huxley formalism, based on a reduction of a leech heart interneuron model, see \cite{Shilnikov2012} and the references therein:
\begin{equation}
\begin{array}{rcl}
C V^\prime &=& -I_{\mathrm{Na}}-I_{\mathrm{K2}}-I_{\mathrm{L}}-I_{\mathrm{app}}-I_{\rm syn} ,  \\
\tau_{\mathrm{Na}}   h^\prime_{\mathrm{Na}} &=& h^\infty_{\mathrm{Na}}(V)-h, \\
\tau_{\mathrm{K2}}   m^\prime_{\mathrm{K2}} &=& m^\infty_{\mathrm{K2}}(V)-m_{\mathrm{K2}}.
\end{array}\label{eq1}
\end{equation}
The dynamics of the above model involve a fast sodium current, $I_{Na}$  with the activation described by the voltage dependent gating variables, $m_{\mathrm{Na}}$ and $h_{\mathrm{Na}}$, a slow potassium current $I_{K2}$ with the inactivation from
$m_{\mathrm{K2}}$, and an ohmic leak current, $I_\mathrm{leak}$:
\begin{equation}
\begin{array}{rcl}
 I_{\mathrm{Na}}&=&{\bar g}_{\mathrm{Na}}\,m^3_{\mathrm{Na}}\,h_{\mathrm{Na}}\,(V-E_{\mathrm{Na}}),\\
 I_{\mathrm{K2}}&=&{\bar g}_{\mathrm{K2}}\,m_{\mathrm{K2}}^2(V-E_{\mathrm{K}}), \\
 I_{L}&=&\bar g_{\mathrm{L}}\,(V-E_{\mathrm{L}}).
 \end{array}
\end{equation}
$C=0.5\mathrm{nF}$ is the membrane capacitance and $I_{\mathrm{app}}=0.006\mathrm{nA}$ is an applied current.
The values of maximal conductances are ${\mathrm{\bar g}}_{\rm K2}=30\mathrm{nS}$, ${\bar g}_{\rm Na}=160\mathrm{nS}$ and
 $\mathrm{g}_{\rm L}=8\mathrm{nS}$. The reversal potentials are $\mathrm{E}_{\rm Na}=45\mathrm{mV}$, $\mathrm{E}_{\rm K}= -70\mathrm{mV}$ and  $E_{\mathrm{L}}=-46\mathrm{mV}$. 

The time constants of gating variables are $\tau_{\rm K2}=0.9\mathrm{s}$ and $\tau_{\rm Na}=0.0405\mathrm{s}$.

The steady state values, $h^\infty_{\mathrm{Na}}(V)$, $m^\infty_{\mathrm{Na}}(V)$, $m^\infty_{\mathrm{K2}}(V)$, of the of gating variables are determined by the following Boltzmann equations:
 \begin{equation}
\begin{array}{rcl}
 h^\infty_{\mathrm{Na}}(V)&=&[1+\exp(500(V+0.0325))]^{-1}\\
 m^\infty_{\mathrm{Na}}(V)&=&[1+\exp(-150(V+0.0305))]^{-1}\\
 \quad m^\infty_{\mathrm{K2}}(V)&=&[1+\exp{(-83(V+0.018+\mathrm{V^{shift}_{K2}}))}]^{-1}.
\end{array}
\end{equation}

Fast, non-delayed synaptic currents in this study are modeled using the fast threshold modulation (FTM) paradigm as follows \cite{ftm}:
 \begin{equation}
 \begin{array}{rcl}
I_{\rm syn} &=& \gsyn (V_{\mathrm{post}}-E_{\rm syn}) \Gamma (V_{\mathrm{pre}}-\Theta_{\rm syn}),\\
\Gamma(V_{\rm pre}-\Theta_{\rm syn}) &=& 1/[1+{\rm exp}\{-1000(V_{\mathrm{pre}}-\Theta_{\rm syn})\}];
\end{array}
\end{equation}
here $V_{\mathrm{post}}$  and $V_{\mathrm{pre}}$ are voltages of the post- and the pre-synaptic cells; the synaptic
threshold $\Theta_{\rm syn}=-0.03\mathrm{V}$ is chosen so that every  spike within a burst in the pre-synaptic cell crosses $\Theta_{\rm syn}$,
see Fig.~\ref{fig1}. This implies that the synaptic current, $I_{\rm syn}$, is initiated as soon as $V_{\mathrm{pre}}$ exceeds the
synaptic threshold. The type, inhibitory or excitatory, of the FTM synapse is determined by the level of the reversal potential, $E_{\rm syn}$,
in the post-synaptic cell.
In the inhibitory case, it is set as $E_{\rm syn}=-0.0625\mathrm{V}$ so that $V_{\rm post}(t)> E_{\rm syn}$. In the excitatory case
the level of $E_{\rm syn}$ is raised to zero to guarantee that the average  of $V_{\rm post}(t)$  over the burst period remains below the reversal potential. We point out that alternative synapse models, such as the alpha and other detailed dynamical representation, do not essentially change the dynamical interactions between these cells \cite{pre2012}.

The numerical simulations and phase analysis were accomplished utilizing the freely available software PyDSTool (version 0.88)  \cite{PyDSTool}. Additional files and instructions are available upon request.

%%%%%%%%%%%%%%%%%%%%

\section*{Acknowledgments} 
J.~S. was supported by the DFG (SCHW 1685/1-1). PhD.studies of J.~W. were funded by NSF grant DMS-1009591. A.~S. was 
is in part supported by NSF grant DMS-1009591, MESRF project 14.740.11.0919, RFFI 11-01-00001 as well as the GSU Brains and Behavior Program.  We would like to thank P. Aswin, S. Jalil, P. Katz, A. Kelley, A. Neiman, W. Kristan,  A. Sakurai, and A. Selverston for stimulating discussions.

% Substitute in the actual, final bib entries!
%\bibliography{ref3_cells}

%%%%%%%%%%%%%%%%%%%%%%%%%%%%%%%%%%%%%%%%%%%%%%%%%%%%%%

\end{document}